\documentclass[aps,pra,twocolumn,floatfix,superscriptaddress,longbibliography]{revtex4-1}
\usepackage{amsmath}
\usepackage{amssymb}
\usepackage{color}
\usepackage{graphics}
\usepackage{graphicx}
\usepackage{dcolumn}
\usepackage{braket}
\usepackage{hyperref}
\hypersetup{
bookmarks=false,
pdftitle={PT-symmetric circuit-QED},
anchorcolor=blue,
colorlinks=true,
citecolor=blue,
urlcolor=blue,
linkcolor=blue}

\begin{document}

\title{$\mathcal {PT}$-symmetric circuit-QED}

\author{Fernando Quijandr\'{\i}a}
\affiliation{Microtechnology and Nanoscience, MC2, Chalmers University of Technology, SE-412 96
G\"oteborg, Sweden}
\author{Uta Naether}
\affiliation{Instituto de Ciencia de Materiales de Aragon and
  Departamento de Fisica de la Materia Condensada, CSIC-Universidad de
  Zaragoza, E-50012 Zaragoza, Spain}
\author{Sahin K. \"Ozdemir}
\affiliation{Department of Engineering Science and Mechanics, The Pennsylvania State University, University Park, PA 16802, USA}
\author{Franco Nori}
\affiliation{CEMS, RIKEN, Saitama 351-0198, Japan}
\affiliation{Department of Physics, University of Michigan, Ann Arbor, Michigan 48109-1040, USA}
\author{David Zueco}
\affiliation{Instituto de Ciencia de Materiales de Aragon and Departamento de Fisica de la Materia Condensada, CSIC-Universidad de Zaragoza, E-50012 Zaragoza, Spain}
\affiliation{Fundacion ARAID, Paseo Maria Agustin 36, E-50004 Zaragoza, Spain}

\begin{abstract}
The Hermiticity axiom of quantum mechanics guarantees that the energy spectrum is real and the time evolution is unitary (probability-preserving).
Nevertheless, non-Hermitian but $\mathcal{PT}$-symmetric Hamiltonians may also have real eigenvalues. Systems described by such effective $\mathcal {PT}$-symmetric Hamiltonians have been realized in experiments using coupled systems with balanced loss (dissipation) and gain (amplification), and their corresponding classical dynamics has been studied. 
A  $\mathcal {PT}$-symmetric system emerging from a quantum dynamics is
highly desirable, in order
to understand what $\mathcal {PT}$-symmetry and the powerful mathematical and physical concepts around it will bring to the next generation of quantum technologies. 
Here, we address this need by proposing and studying a circuit-QED architecture 
that consists
of two coupled resonators and two qubits (each coupled to one resonator). 
By means of external driving fields on the qubits,
we are able to tune gain and losses in the resonators.
Starting with the quantum dynamics of this system, we show the emergence of the
 $\mathcal {PT}$-symmetry via the selection of both driving amplitudes and frequencies. 
 We engineer the system such that a non-number conserving dipole-dipole interaction emerges, introducing an instability at large coupling strengths. The $\mathcal {PT}$-symmetry and its breaking, as well as the predicted instability in this circuit-QED system can be observed in a transmission experiment.
\end{abstract}

\pacs{42.50.Ct, 42.50.Hz, 42.65.-k, 78.20.Bh}

\maketitle


\section{Introduction}

One of the mathematical axioms of quantum mechanics is that the Hamiltonian $H$ of a system should be Hermitian, i.e., $H=H^{\dag}$.
This axiom ensures real energy eigenvalues and, correspondingly, a unitary time evolution, for which the probability to find the system at some state is conserved. Physical systems described by Hermitian Hamiltonians represent closed systems. However,  physical systems in general are open and they are in continuous energy exchange with other systems, experiencing dissipation (or absorption, loss) or receiving energy (gain) from a source. Such systems with gain or loss are described by non-Hermitian Hamiltonians, i.e., $H \neq H^{\dag}$ for which the probability, in general, is not conserved and its time-evolution is not unitary. It is worth pointing out here that open systems  at zero temperature can \emph{effectively} be described by non-Hermitian Hamiltonians.

\begin{figure*}[t]
\begin{center}
\includegraphics[width=1.\columnwidth]{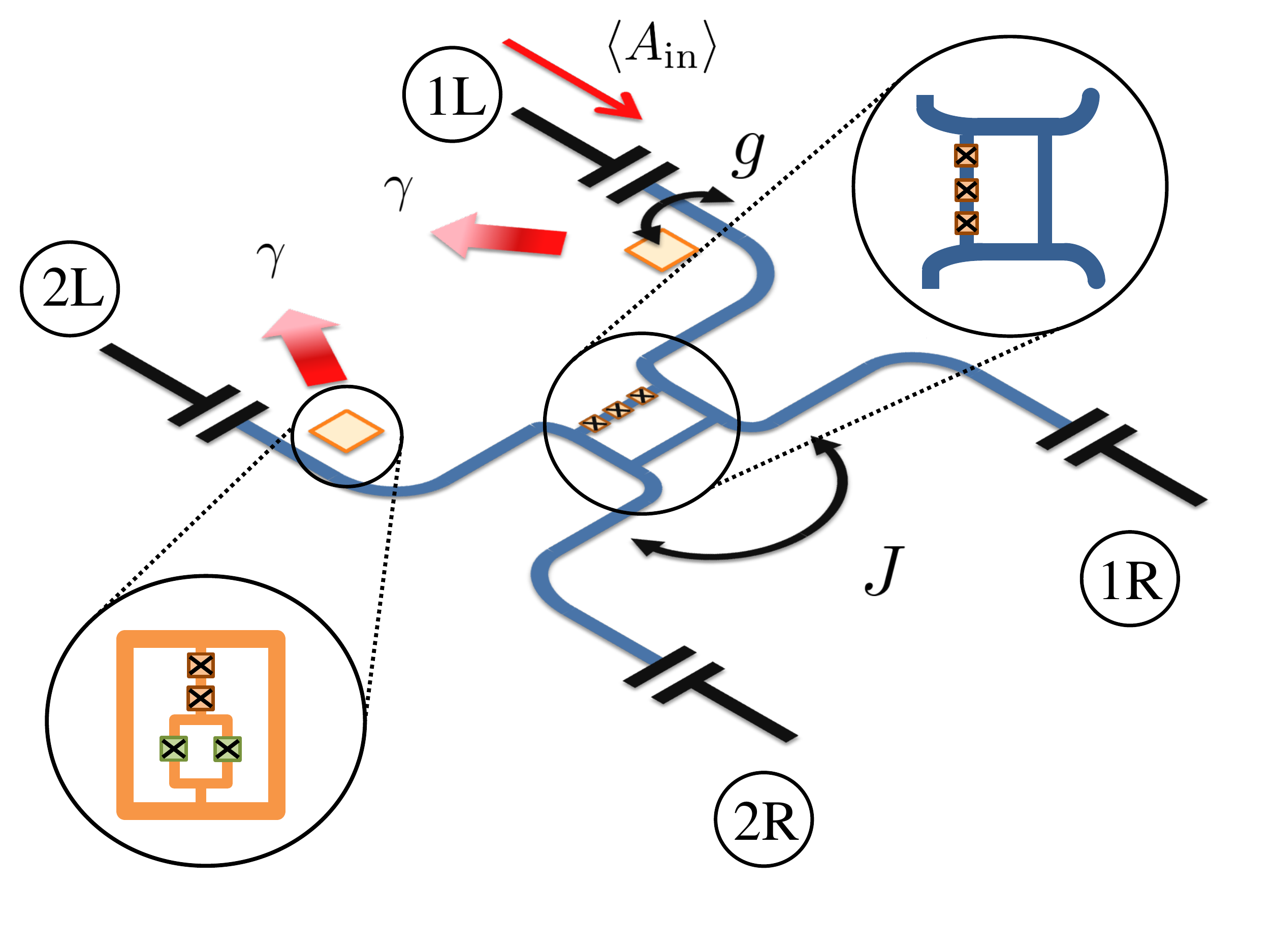}
\end{center}
\caption{A graphical illustration of the proposed circuit-QED architecture to study the physics of $\mathcal {PT}$-symmetry.
Two superconductor resonators are coupled to each other with a coupling strength $J$, and two qubits with decay rates of $\gamma$, each coupled to one resonator with a coupling strength  $g$,  form the basic ingredients of this architecture. The inset in orange color shows the structure of the qubits, whereas the inset in blue shows a possible implementation of tunable coupling between the resonators. }
\label{fig:scheme}
\end{figure*}

In 1998 it was shown \cite{Bender1998}, however, that Hermiticity is not a necessary condition for $H$ to have real eigenvalues.
In fact, a whole class of 
Hamiltonians can have real eigenvalues without being Hermitian, if they are $\mathcal {PT}$-symmetric in the sense that they commute with the $\mathcal {PT}$ operator, \emph{i.e.}, $[\mathcal {PT},H]=0$, where ${\mathcal P}$ is the unitary parity operator and ${\mathcal T}$ is the  anti-unitary time-reversal operator \cite{Bender2005}. It is now understood that one can interpret $\mathcal {PT}$-symmetric systems as non-isolated physical systems having balanced absorption (loss) and amplification (gain). Remarkably, such systems exhibit a phase transition --- spontaneous $\mathcal {PT}$-symmetry breaking ---
at an exceptional point (EP), where both the eigenvalues and the corresponding eigenstates of the system coalesce,
if the parameter that controls the degree of non-Hermiticity exceeds a critical value. Beyond this critical threshold the spectrum is no longer real, and eigenvalues become complex even though $[\mathcal {PT},H]=0$ is still satisfied. In other words, the system experiences a \emph{real-to-complex spectral phase transition}.

The presence of an EP (or a $\mathcal {PT}$ phase transition) drastically affects the dynamics of the system leading to counterintuitive features which can help to control wave transport and light-matter interactions. 
Thus, the field surrounding the concepts of $\mathcal {PT}$-symmetry and EPs (that started as a purely mathematical concept) have turned into a rapidly growing field with many interesting experiments \cite{Ruter2010, Guo2009,Brandstetter2014, Peng2014a, Peng2014, Regensburger2012, Peng2016a, Miao2016, Gao2015, Peng2016b, Fleury2015, Schindler2011, Kim2016, Doppler2016, Chen2017, Hodaei2017, Xu2016, Bender2013, Bittner2012, Chtchelkatchev2012}, 
most of them in the field of optics.
Among the nontrivial phenomena observed in these experiments are  unidirectional invisibility in fiber networks~\cite{Regensburger2012}, nonreciprocal light transport in whispering gallery microresonators \cite{Peng2014}, single-mode lasing in otherwise multimoded lasers with $\mathcal{PT}$-symmetry \cite{Feng2014,Hodaei2014},  loss-induced lasing~\cite{Peng2014a}, control of emission direction of lasing in microring lasers \cite{Peng2016a}, a mobility edge in disordered optical waveguide arrays~\cite{EichelKraut2013,Longhi2015}, as well as chiral dynamics \cite{Doppler2016} and  topological energy transfer when encircling an EP \cite{Xu2016}. Recent years have also seen a number of very interesting theoretical proposals revealing how $\mathcal {PT}$-symmetry can be used to enhance and control optomechanical interactions, and how $\mathcal {PT}$-symmetry affects quantum phase transitions and information retrieval in quantum systems \cite{Kawabata2017, Ashida2017, Gardas2016}. For example, the works of Jing \emph{et al.} with optomechanical microresonators have revealed the possibility of thresholdless phonon lasing \cite{Jing2014}, group velocity control via optomechanically-induced transparency \cite{Jing2015}, enhanced optomechanical cooling at high-order exceptional points \cite{Jing2017}, as well as phonon-analog of loss-induced lasing in optomechanical systems with two-level system defects \cite{Lu2017}. The above mentioned theoretical and experimental works are just a few examples showing the enormous and growing interest in  $\mathcal{PT}$-symmetric systems and their realizations.

In the reported experimental works on $\mathcal {PT}$-symmetry and EPs, open classical systems are engineered such that the dynamics for  the variables of interest obey the $\mathcal {PT}$-symmetry. A study of $\mathcal {PT}$-symmetry and its breaking in experimentally accessible quantum systems is highly desirable to understand the pros and cons of  $\mathcal {PT}$-symmetry for developing quantum technologies. In this work, we address this need by proposing a circuit-QED architecture --- a superconducting circuit operating in the quantum limit \cite{Devoret1995, You2003, Wallraff2004, You2005, You2011, Gu2017}. 
 Starting with a microscopic and unitary description, we demonstrate that the dynamics of this circuit-QED system can be described by an \emph{effective} non-Hermitian $\mathcal {PT}$-symmetric Hamiltonian.

The proposed circuit-QED architecture is experimentally accessible because the main ingredients, a tunable coupling between resonators \cite{Baust2015}, as well as tunability of a qubit gap \cite{You2007b, You2008, Paauw2009, Schwarz2013, You2007, Koch2007}, have already been experimentally demonstrated and are readily available. A circuit-QED architecture for studying $\mathcal {PT}$-symmetry will not only bring the field into the quantum realm but will also offer numerous advantages. For example, so far all $\mathcal {PT}$ symmetry experiments (except Ref.~\cite{Hodaei2017}) involve two components, one with loss and the other with gain. 
These systems are not scalable and 
thus, it is very difficult to expand them
to a larger number of components 
in order
to study collective behavior or the effect of global and local $\mathcal {PT}$-symmetries on wave transport and light matter interactions. Circuit-QED architectures with their scalability (e.g. fabricating arrays of $\mathcal{PT}$-symmetric resonators and coupled qubits with a small footprint should not be a big challenge with current state-of-the-art technologies), and versatility (e.g. engineering  different Hamiltonians by tuning the strength and the frequency of external drives is a natural scenario in circuit-QED) will help to overcome such shortcomings and fabrication difficulties of current platforms used in $\mathcal{PT}$ experiments. Moreover, circuit-QED provides flexibility to explore different parameter regimes which are difficult to reach in current $\mathcal{PT}$ platforms. For instance, 
the ultra- and deep-strong  coupling regimes in resonator-qubit interactions.

This paper is organized as follows. In Section II, we introduce the circuit-QED platform that we propose for the realization of $\mathcal{PT}$-symmetry and its breaking. In Section III we show how one can engineer interactions that either conserve or do not conserve the number of excitations. In Sections IV and V, we derive the effective $\mathcal{PT}$-symmetric Hamiltonian for the system and study its dynamics in the exact and broken $\mathcal{PT}$ phases. In Section VI, we discuss how one can probe the behavior of this circuit-QED platform in the exact and broken $\mathcal{PT}$ phases by transmission experiments. We conclude the manuscript in Section VII by giving a summary of our findings and future prospects. The manuscript is also accompanied by Appendices A and B where details of the derivations are provided.


\section{The circuit}

The circuit-QED architecture we propose for studying $\mathcal{PT}$-symmetry is  sketched in Fig.\ref{fig:scheme}.
It consists of two coupled resonators (blue in the figure) whose coupling can be tuned over time.  An experimental demonstration of a tunable coupling through a
three-junction loop (sketched in the center of  figure \ref{fig:scheme} and zoomed in the right top corner) was recently reported \cite{Baust2015, Baust2016}. Each resonator is coupled to one qubit (orange boxes in Fig. \ref{fig:scheme}) that has a tunable gap (e.g., flux qubits \cite{Paauw2009, Schwarz2013} or capacitively shunted qubits \cite{You2007, Koch2007}. The system is described by the Hamiltonian
\begin{equation}
\label{Hmicro}
H(t) = H_0(t) + H_c(t) ,
\end{equation}
where
\begin{equation}
\label{H0}
H_0(t) = \sum_{j=1,2}
\omega_j a_j^\dagger a_j
+
\frac{\epsilon_j(t) }{2} \sigma_j^z \,,
\end{equation}
represents the free part of the Hamiltonian. Here, $\omega_j$ are the bare frequencies of the resonators and $\epsilon_j (t)$ represent the qubit gaps that can be tuned in time. These building blocks are coupled via the interaction Hamiltonian
\begin{equation}
\label{Hc}
H_c(t)=\sum_{j=1,2} g_j \sigma_j^x (a_j +a_j^\dagger)
+
J(t) (a_1^{\dagger} + a_1)( a_2^{\dagger} + a_2) \,,
\end{equation}
with coupling strengths $g_j$ and $J(t)$. The time tunability of the gaps $\epsilon_j (t)$ and the resonator-resonator coupling $J(t)$ turns to be crucial in what follows.

Apart from the unitary evolution governed by $H(t)$, both qubits and resonators are coupled to the circuitry environment. The influence of the latter in circuit QED is weak (compared to the order of the bare system frequencies) and therefore, it suffices to treat it with a master equation of the optical type
\begin{equation}
\label{qmefull}
\frac{{\rm d}}{{\rm d}t} \varrho = - i [H(t), \varrho] +  \sum_j \gamma_j {\mathcal
  D} [\sigma_j ] \varrho +
 \kappa_j {\mathcal D} [a_j] \varrho ,
\end{equation}
where $\gamma_j^{-1} \; (\kappa_j^{-1})$ accounts for the time scale of relaxation to
equilibrium ($\gamma_j^{-1} \sim T_j$)  driven by  the dissipators  ${\mathcal D} [o_n] \varrho = o_n  \varrho
o_n^\dagger - \frac{1}{2} ( o_n^\dagger o_n \varrho +
\varrho o_n^\dagger o_n ) $.



\section{Hamiltonian engineering}\label{sec:engine}

In the following, we will work in the interaction picture with respect  to $H_0 (t)$, and assume that the qubit-gap is modulated as,
\begin{equation}
\label{drive}
\epsilon_j (t) = \epsilon_j^{(0)} +
\sum_{\alpha = \pm}
 \lambda_{j, \alpha} \cos ( \Omega_{j, \alpha} t) ,
\end{equation}
for $j = 1,\,2$, where $\Omega_{j, \pm}$ represent the  driving frequencies given by
\begin{equation}
\label{Omegas}
\Omega_{j, \pm} = \epsilon_j^{(0)} \pm (\omega_j + \delta).
\end{equation}
In order to validate our approximations, we are going to restrict ourselves to the following
hierarchy in parameter space:
\begin{equation}
\label{hiefull}
\epsilon_j^{(0)} \gg \gamma_j \gg \omega_j \gg \, g_j, \,
J, \, \delta \, \text{; and }\Omega_{j, \pm}> \lambda_{j , \pm} \, .
\end{equation}
Finally,  we assume that $\omega_1 = \omega_2 = \omega$, which is more than plausible due to well-established and highly reproducible fabrication 
of superconducting resonators \cite{Underwood2012}. In our numerical tests we  set $\epsilon_j \cong 5 \omega$, $\gamma \cong 2
\omega$, and $g \sim J \cong 10^{-2} \omega$.  These parameters are reasonable from the experimental point of view and serve to justify all the approximations made below.

\begin{figure}[t]
\begin{center}
\includegraphics[width=1.0\columnwidth]{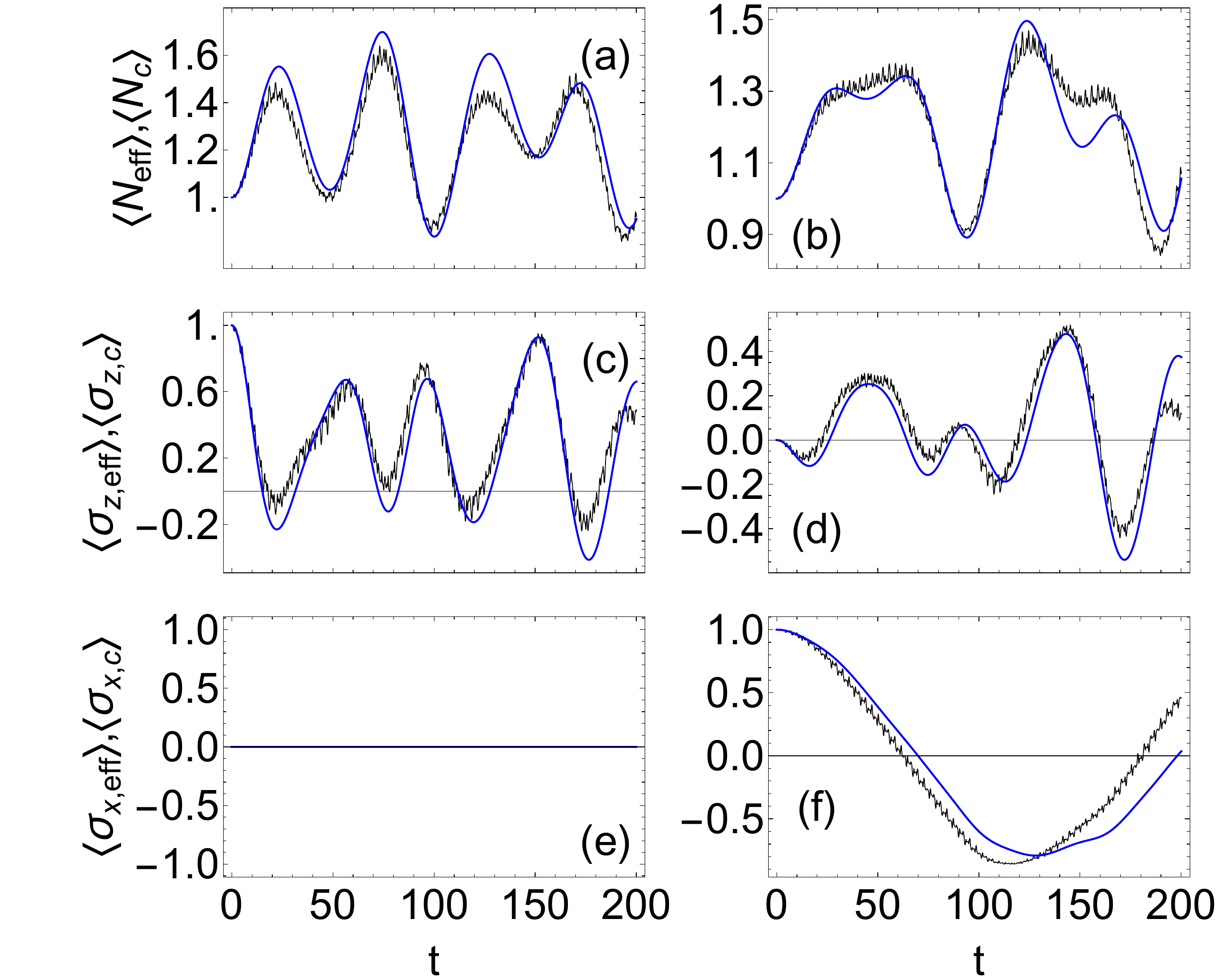}
\end{center}
\caption{Time evolution of $\langle N\rangle$ [in (a) and (b)], $\langle \sigma_z\rangle$ [in (c) and (d)] and $\langle \sigma_x\rangle$ [in (e) and (f)] in the proposed circuit-QED architecture according to Eqs. \eqref{Hctime} (black) and  \eqref{appHeffRWA} (blue). The parameters used in the numerical simulations are:
 $\lambda_{+}=\lambda_{-}=2$, $\epsilon=5$, $\omega=1$, $\delta=0.1$, $g=0.05$, $G_-/G_+=0.58$, for the initial state [(a), (c) and (e)]: $|\psi(0)\rangle=|\uparrow\rangle|1\rangle$.
and for the initial state [(b), (d) and (f)]:  $|\psi(0)\rangle=(|\uparrow\rangle+|\downarrow\rangle)|1\rangle/\sqrt{2}$.
}
\label{fig:engin}
\end{figure}

\begin{figure}[t]
\begin{center}
\includegraphics[width=1.0\columnwidth]{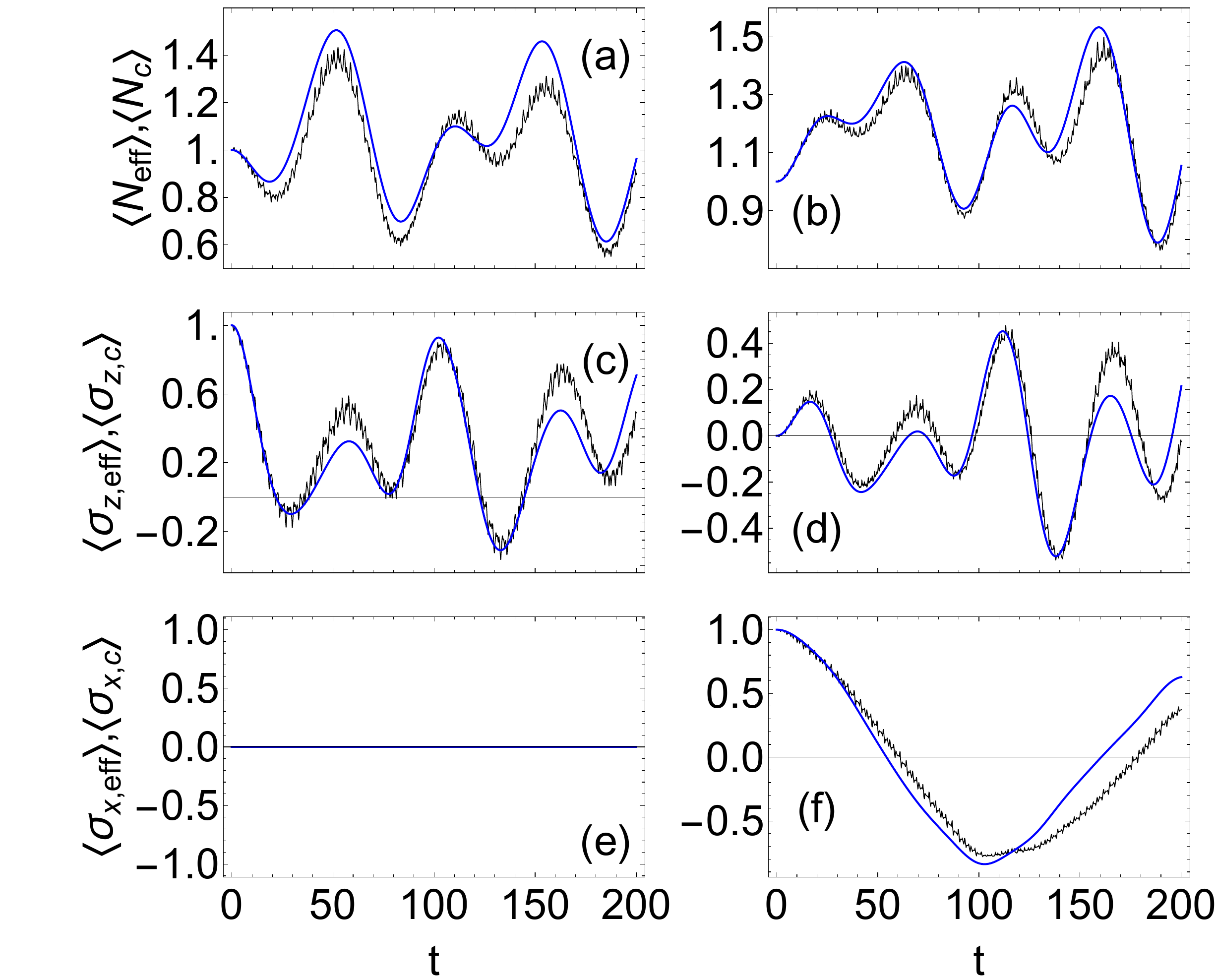}
\end{center}
\caption{Time evolution of $\langle N\rangle$ [in (a) and (b)], $\langle \sigma_z\rangle$ [in (c) and (d)] and $\langle \sigma_x\rangle$ [in (e) and (f)] in the proposed circuit-QED architecture according to Eqs. \eqref{Hctime} (black) and  \eqref{Hceff} (blue). The parameters used in the numerical simulations are:
 $\lambda_{+}=3, \lambda_{-}=1$, $\epsilon=5$, $\omega=1$, $\delta=0.1$, $g=0.05$, $G_-/G_+=2.1$, for the initial state [(a), (c) and (e)]: $|\psi(0)\rangle=|\uparrow\rangle|1\rangle$.
and for the initial state [(b), (d) and (f)]:  $|\psi(0)\rangle=(|\uparrow\rangle+|\downarrow\rangle)|1\rangle/\sqrt{2}$.
}
\label{fig:engin2}
\end{figure}

The interaction Hamiltonian in the interaction picture with respect to $H_0(t) $  [Cf. Eqs. \eqref{Hmicro}, \eqref{H0}, \eqref{Hc} and \eqref{drive}] is
\begin{align}\label{Hctime}
\widetilde{H}_c(t)  &=  J(a_1^{\dagger} a_2{\rm e}^{i (\omega_1 - \omega_2) t} +
a_1 a_2 \,{\rm e}^{-i (\omega_1+\omega_2)t} )
\\ \nonumber
&+ g\sigma^x_1(t) a_1 {\rm e}^{-i\omega_1 t}
+ g\sigma^x_2(t) a_2 {\rm e}^{-i\omega_2 t} + {\rm h.c.}
\end{align}
where
\begin{align}
\sigma^x_{j}(t)  & = \sigma^+_j  \, \exp \left( i\epsilon_j^{(0)} t + \sum_\alpha 2i\frac{\lambda_{j,\alpha}}{\Omega_{j, \alpha}} \sin (\Omega_{j, \alpha} t) \right) +
{\rm h.c.}
\nonumber\\
& = \sigma^+_j  {\rm
  e}^{ i\epsilon_j^{(0)} t}
\prod_\alpha \sum _n   J_n \left(\frac{2\lambda_{j,\alpha}}{\Omega_{j,\alpha} }\right) \, {\rm
  e}^{ i t n \Omega_{j, \alpha} }
+
{\rm h.c.}
\end{align}
with $J_n$ representing the $n$-th Bessel function of the first kind. By choosing $\Omega_{j,\pm}$ according to \eqref{Omegas}
and recalling the \emph{necessary hierarchy} \eqref{hiefull}, the Hamiltonian \eqref{Hctime} can be approximated (neglecting terms
oscillating with $\epsilon_j^{(0)}$) as
\begin{align}
\label{Hceff}
\widetilde{H}_c(t)  & \cong J \Big (a_1^{\dagger} a_2 {\rm e}^{i (\omega_1 - \omega_2) t} +
a_1 a_2 \,{\rm e}^{-i (\omega_1+\omega_2)t} \Big  )
\\ \nonumber
&+
\sum_j g_j \Big ( G_{j +} \sigma^+_j a_j {\rm e}^{i \delta t} + G_{j -}
\sigma^+_j a^{\dagger}_j {\rm e}^{-i \delta t} \Big )  + {\rm h.c.}
\end{align}
with $G_{j, \pm}$ given as
\begin{equation}
\label{Gjpm}
G_{j, \pm} = J_0 \left( 2 \frac{\lambda_{j, \pm} }{ \Omega_{j, \pm} } \right) J_1 \left( 2 \frac{ \lambda_{j, \mp} }{ \Omega_{j, \mp} } \right) \,.
\end{equation}


\subsection{Engineering a number-conserving interaction}

If the resonator-resonator coupling $J$  is constant, the second term inside the first parenthesis of \eqref{Hceff} can be neglected following the hierarchy (\ref{hiefull}).
Here, we assume that $\omega_1 = \omega_2$. In order to get rid of the extra time dependence (due to $\delta$), we move to a frame rotating with this frequency. Then, the effective Hamiltonian can be written as:
\begin{align}
\label{appHeffRWA}
H_{\rm eff}'  & \cong  - \sum_j \delta a_j^\dagger a_j +
               J(a_1^{\dagger} a_2 +{\rm h.c.} )
\nonumber\\
&+
\sum_j g_j \left[ \Big ( G_{j +} \sigma^+_j a_j + G_{j -}
\sigma^+_j a_j^{\dagger} \Big )  + {\rm h.c.}
\right] .
\end{align}
The validity of the approximations (following the hierarchy \eqref{hiefull}) was tested and the results are shown in Figs.~\ref{fig:engin} and \ref{fig:engin2} for a single
resonator coupled to a qubit. There we show the time evolution under $\widetilde{H}_c(t)$ \eqref{Hctime} in black and $H_{\rm eff}'$
 \eqref{appHeffRWA} in blue. Initial states are  $\left|
  \phi(t=0)\right.\rangle=\left| \uparrow\right.\rangle\left|
  1\right.\rangle$ (left) and $\left|
  \phi(0)\right.\rangle=\frac{1}{\sqrt{2}}\left[\left|
    \uparrow\right.\rangle\left| 1\right.\rangle+\left|
    \downarrow\right.\rangle\left| 1\right.\rangle\right]$
(right). Here $| \downarrow\rangle$ and $| \uparrow\rangle$ are the ground and excited states of the qubit, respectively, and $| n \rangle$
 are the Fock states. We compare the time evolution of \emph{one}  resonator  of frequency $\omega=1$ coupled to \emph{one} qubit driven with frequencies $\Omega_+$ and $\Omega_-$, assuming $\delta=0.1$. The driving amplitudes $\lambda_+=\lambda_-$ in 
 Fig.~ \ref{fig:engin} are chosen equal, giving a ratio $G_{-}/G_{+}<1$. In Fig.~ \ref{fig:engin2}, we choose the  $\lambda_{\pm}$ in such a way that $G_{-}/G_{+}>1$. 
 It is seen that our approach holds for both the loss- and gain-dominant cases. The fluctuations of the time-dependent Hamiltonian \eqref{Hctime} are relatively small and can be made even smaller by decreasing $g$ or increasing $\epsilon$. This is shown on the left-hand-side of 
 Figs. ~\ref{fig:engin} and \ref{fig:engin2}, where the values of $g$ and $\epsilon$ are smaller and larger, respectively, than their counterparts on the right-hand-side.


\subsection{The non-conserving number case}
\label{sec:JnRWA}

Here we want to exploit the possibility of an on-time tuning of the resonator-resonator coupling \cite{Peropadre2012,Felicetti2014}. By setting 
\begin{equation}
J (t) = J \Big ( \cos  [ (\omega_1 + \omega_2+ 2 \delta ) t] +
\cos  [ (\omega_1 - \omega_2 ) t]  \Big )
\;.
\end{equation}
the effective Hamiltonian becomes
\begin{align}
\label{Heffus}
H_{\rm eff}  &\equiv H_{\delta,J} + H_{g_j}^{\rm eff}
\nonumber\\ &\cong  - \sum_j \delta a_j^\dagger a_j +
                          J(a_1^{\dagger} a_2 + a_1 a_2 )
\nonumber\\
& +
\sum_j g_j \left[ \Big ( G_{j +} \sigma^+_j a_j + G_{j -}
\sigma^+_j a_j^{\dagger}  \Big )+ {\rm h.c.} \right] \,
\end{align}
where the coupling between resonators includes both the number-preserving terms $a^\dagger_1 a_2 + a_1 a^\dagger_2$, as
well as the counter-rotating terms $a^\dagger_1 a^\dagger_2 + a_1 a_2$. Recall that within the Rotating Wave Approximation (RWA), the counter-rotating terms are neglected [Cf. Eq. \eqref{appHeffRWA}]. 
Therefore, we have an \emph{effective} model which allows us to study physics (beyond the RWA) in the so-called ultra- and deep-strong coupling regimes (borrowing the nomenclature from the light-matter Rabi model). In our case, it is not the actual coupling strength but the time dependence, $J(t)$,  or the ratio $J/\delta$ which sets if the RWA is valid or not. The scenario described here serves as a controllable example where RWA versus non-RWA physics may be investigated.

In the following, we will concentrate in the dynamics governed by Eq.~\eqref{Heffus} but we will compare it with the number-conserving case given in Eq.~\eqref{appHeffRWA}, which is the one mainly studied in the literature.



\section{Effective $\mathcal{PT}$ equations }

\subsection{Adiabatic elimination}

Let us now deal with the dissipative part of Eq. ~(\ref{qmefull}).  The effective time scales in \eqref{Heffus} are given by $ \delta , \;J$ and  $ g_j$.  In the range
defined by Eq. ~(\ref{hiefull}),  the fastest dynamics corresponds to the dissipative evolution of the (bad) qubits: $\gamma_j$.  In this regime, we can  adiabatically eliminate the
qubit's degrees of freedom. In doing so,  we end up with the slow part, which solely describes an effective dynamics for the two resonators. The technicalities of the adiabatic elimination were discussed already in Ref. \cite{Cirac1992} and adapted for a similar setup in Ref. \cite{Quijandria2013}. In Appendix \ref{app:adia} we give the details of lengthy manipulations, and here we prefer to write directly the effective equations for the first moments of the resonators' operators after eliminating the qubit degrees of freedom:
\begin{equation}
\label{aeqio}
\frac{{\rm d}}{{\rm d} t} \langle  a_j  \rangle
=
i \langle [H_{\rm\delta,J} , a_{j}] \rangle
+ \sum_j \frac{2 g_j^2}{\gamma_j}
\langle D^\dagger [b_j] a_j \rangle.
\end{equation}
The $b_j$ operators are defined as
[Cf. Eq. \eqref{Gjpm}]
\begin{equation}
b_1 = G_{1, +}  \, a_1 +  G_{1,-} \,  a_1^{\dagger}
\qquad
b_2 = G_{2,+}  \,  a_2 + G_{2,-}  \,  a_2^{\dagger}
\; .
\end{equation}

\begin{figure}
\begin{center}
\includegraphics[width=0.90\columnwidth]{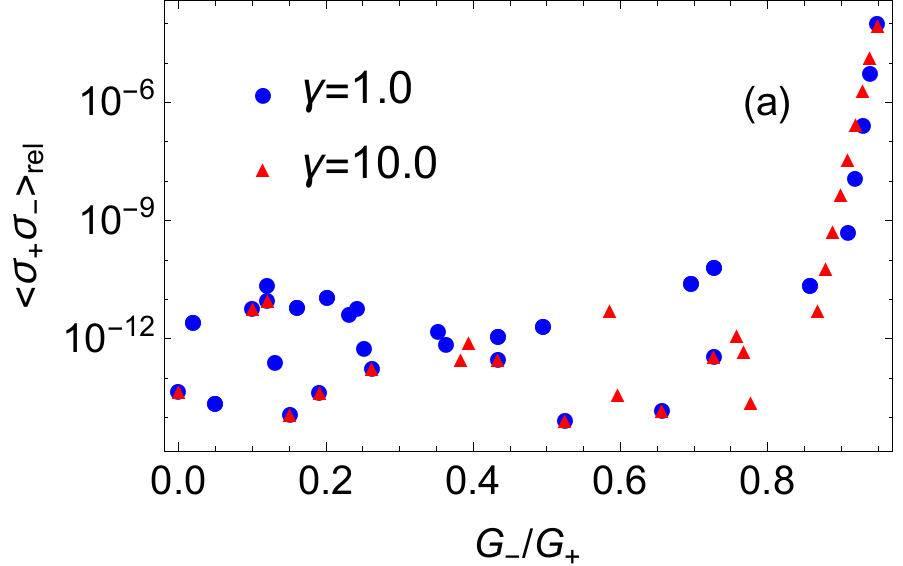}
\includegraphics[width=0.90\columnwidth]{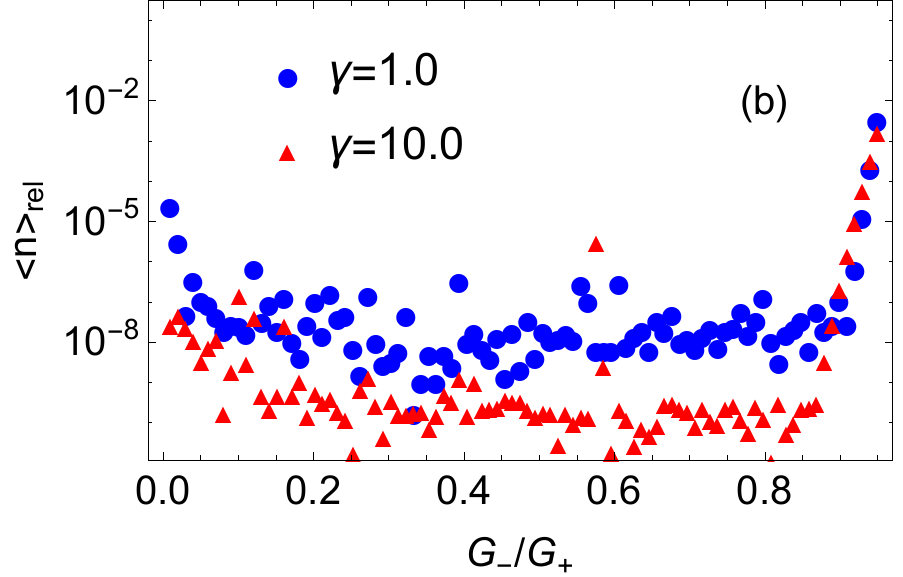}
\end{center}
\caption{ Numerical verification of the adiabatic elimination for
a single resonator coupled to a qubit with decay rates of $\gamma=1$ and $\gamma=10$. We use a Fock space of dimension
$N_{\rm Fock}=300$.
In (a) we plot the relative error for $\sigma^+ \sigma^-$
as a function of  $G_{-}/G_{+}$. In (b) we plot the relative error for   $n=a^{\dagger} a$.
 }
\label{fig:adiabatic}
\end{figure}

In order to test the validity of the adiabatic elimination, we compare the \emph{stationary}
values obtained from \eqref{aeqio} with those obtained from the full quantum master
equation given in Eq. \eqref{qmefull}. In Fig. \ref{fig:adiabatic} we plot the relative error between the results obtained from
both equations for the stationary values of $\langle\sigma^+ \sigma^-\rangle$ and $\langle a^\dagger a \rangle$. We do it for the case of a single resonator coupled to a qubit and as a function of the ratio $G_{-} / G_{+}$. This is the squeezing parameter in the dissipative dynamics of \eqref{aeqio} which fully determines the stationary solutions \cite{Quijandria2015}. Our numerical results support the validity of the adiabatic elimination.

\subsection{Verifying the symmetries}

To write Eq. ~(\ref{aeqio}) in a more convenient way,  we define  the vector $\vec \alpha ^t := ( \langle a_1 \rangle, \langle
a_1^\dagger \rangle, \langle a_2 \rangle, \langle a_2^\dagger \rangle
)$, as well as the effective decay rates
\begin{equation}
\label{redefgam}
\widetilde \gamma_j := (-1)^{j+1} \frac{2 g_j^2}{\gamma_j} (G_{j, -}^2
- G_{j,+}^2 ) \, .
\end{equation}
Note the $(-1)^{j+1}$ prefactor in the above equation. By setting  $G_{1,-}> G_{1,+}$ and  $G_{2,-} <
G_{2,+}$  we have $\widetilde \gamma_j >0$ always. These relations among the $G's$ imply that the resonator $1$ is dissipating (i.e., losses are larger than the gain) and the resonator $2$ is amplifying (i.e., gain is larger than losses).  In doing so, Eq. \eqref{aeqio} defines the set:
\begin{equation}
\label{alphaM}
i \frac{{\rm d}}{{\rm d} t} \vec \alpha = M \vec \alpha
\end{equation}
with,
\begin{equation}
\label{M}
M =
\begin{pmatrix}
  \delta -i \widetilde \gamma_1 &0 & J &   \, J  \\
  0 & -\delta-i \widetilde\gamma_1 &
-  \, J & -J \\
   J&  \,J  & \delta +i \widetilde\gamma_2 & 0  \\
  - \,J & - J & 0 & - \delta + i \widetilde\gamma_2
 \end{pmatrix} \,.
\end{equation}
By representing  the \emph {unitary} parity operator  by
\begin{equation}
\mathcal {P} M \mathcal {P} = (\sigma_x \otimes {\mathbb I}_2)  \, M \,
(\sigma_x \otimes {\mathbb I}_2),
\end{equation}
and the  \emph {anti-unitary}  time reversal one by
\begin{equation}
\mathcal {T} M \mathcal {T} = M^*
\end{equation}
one can directly verify that in the balanced gain-loss case,
$\widetilde \gamma_1 =\widetilde \gamma_2 = \widetilde \gamma$, the matrix $M$ is $\mathcal{PT} $-symmetric, i.e.,
$[\mathcal{PT}, M]=0$.


\section{Broken $\mathcal {PT}$-symmetry \textcolor[rgb]{1.00,0.00,0.00}{Phase}}

\begin{figure}[t]
\begin{center}
\includegraphics[width=0.95\columnwidth]{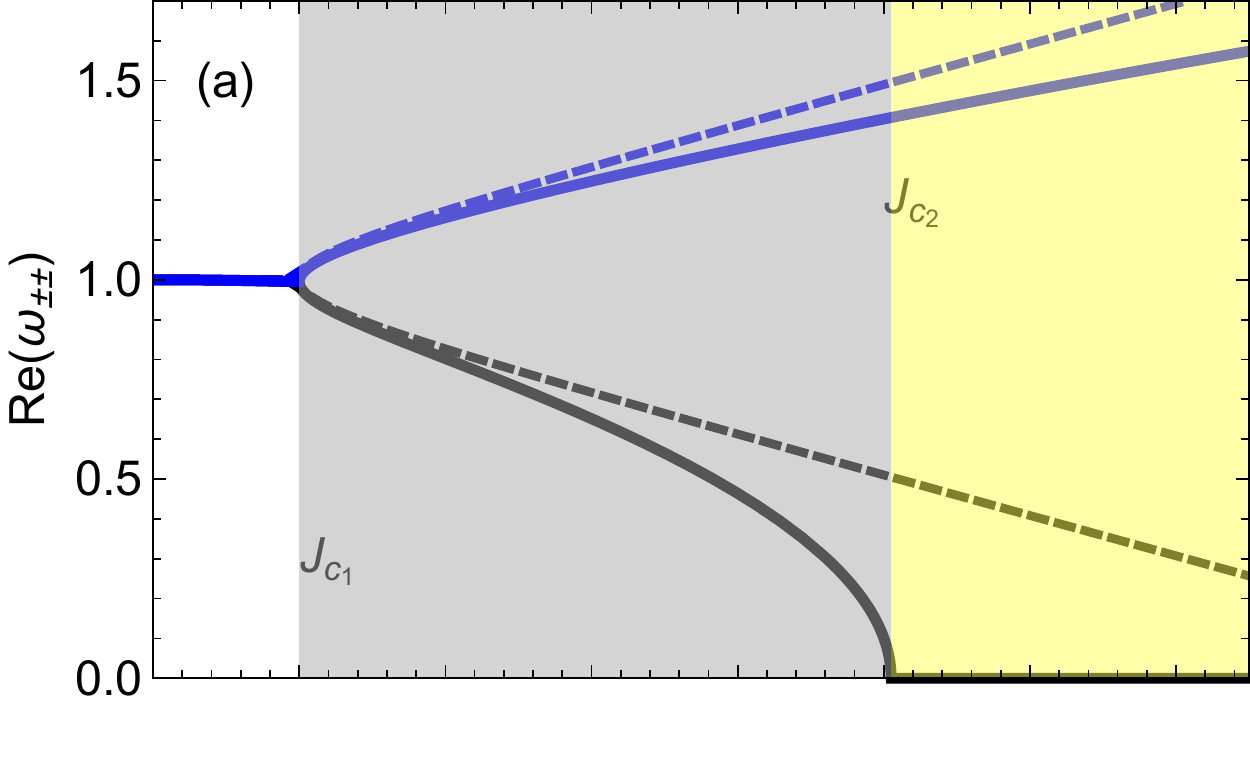}
\includegraphics[width=0.95\columnwidth]{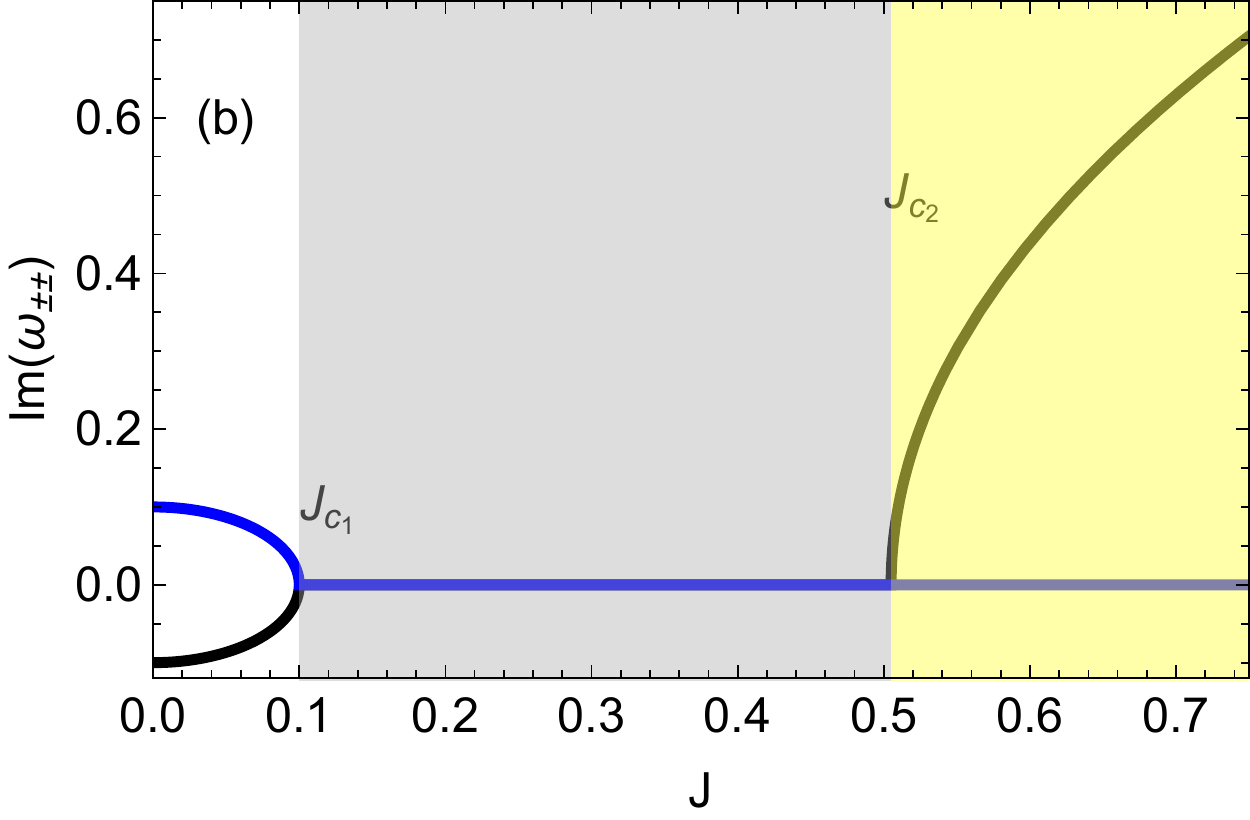}
\end{center}
\caption{Evolution of the eigenvalues of the proposed circuit-QED architecture as a function of the resonator-resonator coupling strength $J$. The background colors distinguish among three regions: $J<J_{c_1}$ (white),   $J_{c_1}<J<J_{c_2}$ (light gray), and $J_{c_2}<J$ (light yellow). The eigenfrequency  $\omega_{+-}$ is plotted in  black while $\omega_{++}$ in  blue [Cf. Eq. \eqref{omegas}]. The dashed lines correspond to the RWA \eqref{appHeffRWA} while the solid lines correspond to the most general case \eqref{Heffus}. In (a) we plot the real part of $\omega_{+ \pm}$ and in (b) the
imaginary part of the eigenvalues.  The parameters used in the simulations are: $\widetilde \gamma_1=\widetilde \gamma_2=0.1$ and $\delta= 1$.
}
\label{fig:ucs}
\end{figure}

As stated in the introduction, $\mathcal {PT}$-symmetric Hamiltonians may exhibit a real spectrum for certain parameter combinations. The (phase) transition from a complex to
a real-valued spectrum occurs at a so-called exceptional point (EP), which marks the degeneracy of a non-Hermitian system, including $\mathcal {PT}$-symmetric systems. At an EP both the eigenvalues and the corresponding eigenvectors of the Hamiltonian coalesce (i.e., become degenerate). Consequently, the non-Hermitian Hamiltonian governing the system becomes non-diagonalizable. This is significantly different than eigenvalue-degeneracies of Hermitian systems where one can always assign orthogonal eigenvectors to degenerate eigenvalues.

In the case of balanced gain and loss, we expect the matrix $M$ to have real eigenvalues in the exact  $\mathcal{PT} $-symmetric phase and complex conjugate eigenvalue pairs in the broken $\mathcal{PT} $-symmetric phase. Diagonalizing  Eq. \eqref{M}  we obtain
\begin{align}
\label{omegas}
\omega_{ \pm \pm} =
\pm
[
\delta^2  -\widetilde \gamma^2
\pm 2 \delta (J^2 - \widetilde \gamma^2)^{1/2} ]^{1/2}
\, .
\end{align}
By simple inspection, one can immediately see that for $J^2 - \widetilde \gamma^2<0$, the eigenvalues expressed in Eq. \eqref{omegas}  are complex, i.e. the system is in the broken $\mathcal{PT} $ phase. We note that even in this phase $[\mathcal{PT}, M]=0$ still holds. The eigenvalues of $M$ are real
whenever
\begin{equation}
\label{Jc1}
J > J_{c_1} = \widetilde \gamma
\, 
\end{equation}
where $J_{c_1}$ corresponds to the $\mathcal{PT}$-transition point (i.e., \emph{real-to-complex spectral phase transition} point) typically observed in experiments (e.g., in \cite{Peng2014,Peng2014a,Peng2014b}).
Figure \ref{fig:ucs} shows the evolution of the eigenvalues given in Eq. \eqref{omegas} as a function of the coupling strength $J$.
Here, the dashed lines correspond to the RWA in the resonator-resonator coupling. 
It is clearly seen that in the RWA model there is only one transition point located at  $J=J_{c_1}$ where the spectra transits from complex to real eigenvalues. 
In agreement with our discussion of EP's, we notice that at $J=J_{c_1}$, both eigenvalues coincide.
In Fig. ~\ref{fig:ucs}, we can also verify that for the general model given in Eq. \eqref{Heffus} (which is beyond the RWA model), there is a second transition at $J=J_{c_2}$ beyond which the eigenvalues become complex again. Thus, in our circuit-QED architecture, real eigenvalues are obtained in the parameter space defined by \begin{equation}
\label{Jc2}
\widetilde \gamma \leq J_{c_1}< J < J_{c_2} = \frac{\widetilde \gamma^2 + \delta ^2}{ 2 \delta }
\, .
\end{equation}
We note that the inequality on the right hand side of Eq. \eqref{Jc2}, which gives the maximum value of $J$ for real eigenvalues, is reported for the first time here. However, we must be cautious in associating this second transition to a breaking of the $\mathcal {PT}$-symmetry.  This is because the  effective Hamiltonian $H_{\delta, J}$ is not positive definite. Therefore, in the absence of dissipation, the eigenfrequencies $\omega_{\pm-}$  are only real for $J < \delta /2$. This bound corresponds to Eq. \eqref{Jc2} for $\widetilde \gamma = 0$. In the dissipative case,  the fact that the eigenvalue $\omega_{+-}$ (solid black line in Fig. \ref{fig:ucs})
becomes complex for $J > J_{c_2}$ (blue region) is a reminiscence of the latter.  Therefore, this second transition should be understood as an instability point of the driven dissipative system.
Another argument to support our claim is that $J = J_{c_2}$ is not an EP.

In Appendix \ref{app:ibev} we give the general expression for the eigenvalues of the matrix $M$ when the gain and loss are not balanced ($\widetilde \gamma_1 \neq\widetilde
\gamma_2$). In this case, there is \emph{always} a non-zero imaginary contribution to the normal frequencies [Cf. Eq. \eqref{omegas_imb}]. Apart from this offset, the 
transitions discussed above can also be traced (see Fig. \ref{fig:imbalanced}).




\section{INPUT-OUTPUT:  TRANSMISSION EXPERIMENTS}

\begin{figure}
\begin{center}
\includegraphics[width=0.48\textwidth]{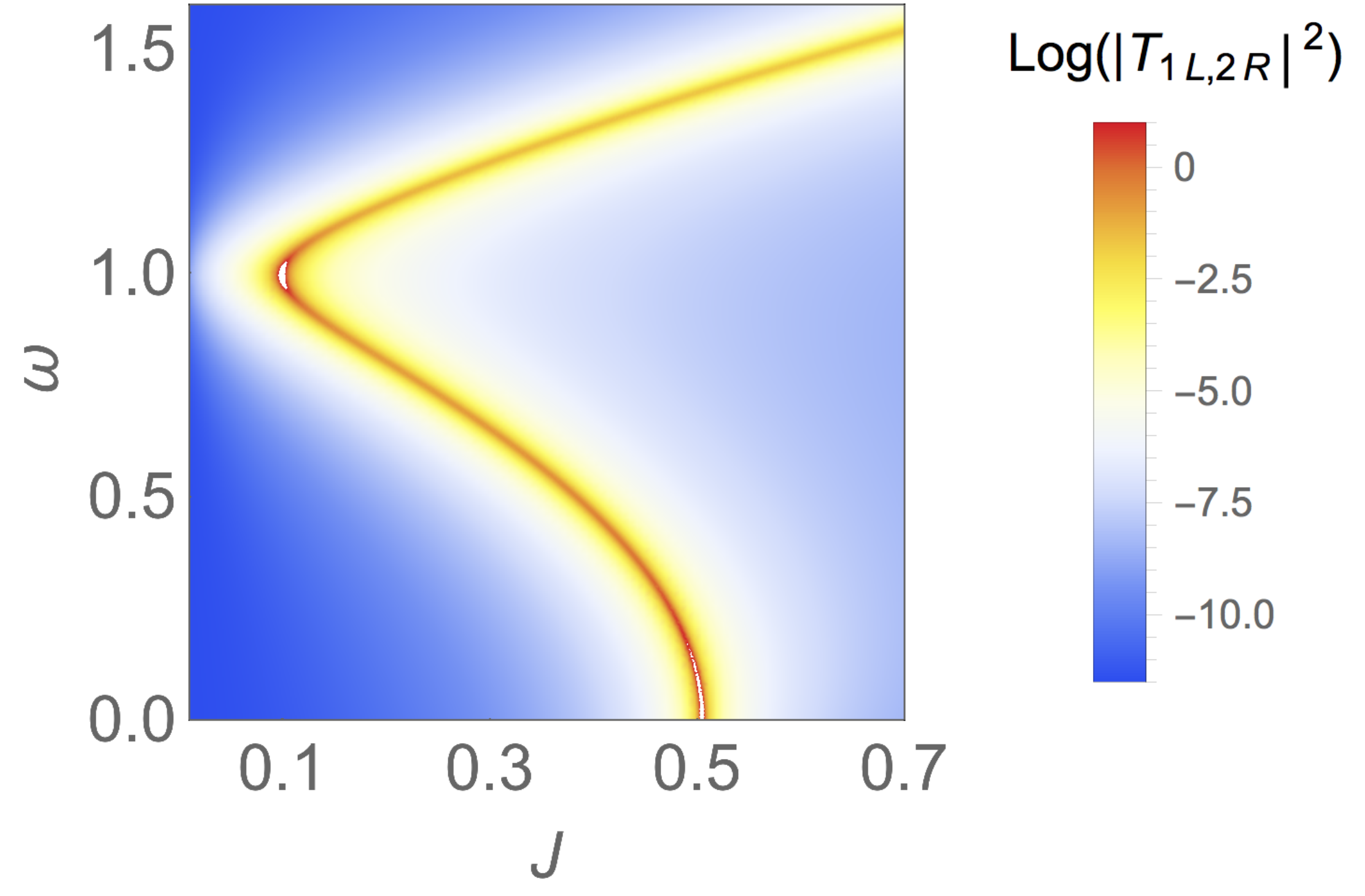}
\end{center}
\caption{
  Logarithm of the power transmitted from port $1L$ to $2R$: $\log(|T_{1L, 2R}|^2)$ as a function of the input frequency $\omega$ and resonantor-resonator coupling $J$.
Parameters used in the simulation are: $\tilde{\gamma}_1 = \tilde{\gamma}_2 = 0.1$ and $\kappa=0.02$. }
\label{fig:ucs_trans}
\end{figure}

In the proposed circuit-QED setup,  the $\mathcal {PT}$-symmetry can be probed with a simple transmission experiment.
Typically, a low-power coherent input $\langle A^{\rm in} \rangle$ is sent into one of the resonators, as depicted in Fig.
\ref{fig:scheme}.  We label the four ports as $1L$, $1R$, $2L$, and $2R$, where the number indicates resonator $1$ or $2$ to which the fields are
coupled to, and the letter $R$  (or $L$) points out whether the field enters or leaves the circuit from the right  (or left)  (See Fig.~\ref{fig:scheme}). In
the figure, the input is sent through $1L$. 
The transmitted
signals (emerging at $1R, 2L, 2R$) or the reflected one (emerging at $1L$) can
be measured with a vector network analyzer.
Indeed, a two-resonator architecture has already been experimentally studied using transmission experiments \cite{Baust2015}. Thus, the same techniques can be directly used for the experimental realization of our proposal. Such experiments are described by the input-output
theory \cite{Gardiner1985}.  The system of interest, in our case the two resonator - two qubit layout, is coupled to external leads (open transmission lines).  We treat both  the system and the signals (input and output)  quantum mechanically.  The fields in the leads are assumed to be bosonic free fields given by
\begin{equation}
H_{\rm leads}=\sum_{j, \lambda} \int {\rm d} \omega A^\dagger_{j,
  \lambda} (\omega) A_{j,
  \lambda} (\omega) \; ,
\end{equation}
 where $[A_{j, \lambda} (\omega), A_{j^\prime, \lambda^\prime}
(\omega^\prime)^\dagger ] = \delta_{j, j^\prime} \, \delta_{\lambda,
  \lambda^\prime} \, \delta (\omega - \omega^\prime)$, with $j=1,2$ and $\lambda=L,R$.
Note that the leads  act  as extra baths for the resonators (adding leakage to the system).

The interaction between the system and the transmission lines (in the case of the proposed setup, a capacitive interaction) is described by
the Hamiltonian
\begin{equation}
H_{\rm int} = \sum _{j, \lambda} \int {\rm d}\omega\, \kappa (\omega) \Big ( a_j   A^\dagger_{j,
  \lambda} (\omega) +   {\rm h.c.} \Big ) \; .
\end{equation}
To obtain the relation for the input and output fields, the Heisenberg equations for the fields $A^\dagger_{j, \lambda} $ are considered and Fourier transformed.  The input fields, defined as $A^{\rm in}_{j, \lambda}(t) = \int_0^\infty {\rm
  d}\omega / \sqrt{2 \pi} \; A_{j, \lambda} (\omega, t_0)  \, {\rm e}^{-i \omega
  t}$
take into account contributions from the leads from a time $t_0$, before the interaction
between the input and the system actually occurs. On the other hand, the output fields   $A^{\rm out}_{j, \lambda} (t) = \int_0^\infty {\rm
  d}\omega / \sqrt{2 \pi} \; A_{j, \lambda} (\omega, t_f)  \, {\rm e}^{-i \omega
  t}$ consider contributions up to a time $t_f$ after the interaction took place.  Without loss of generality, a monochromatic signal $\langle A^{\rm in}_{1L} \rangle= \alpha {\rm
  e}^{i \omega_d t}$ can be used.
Following Ref. \cite{Gardiner1985}, the input-output relation is
\begin{equation}
\label{in-out}
\langle A^{\rm out}_{j, \lambda} \rangle = \langle  A^{\rm in}_{j, \lambda}  \rangle
 -i \sqrt{K} \;
\chi_{a_j}^{ A^{\rm in}} \,,
\qquad 
j=1,2 \,; \lambda=L, R
\end{equation}
where $ K = 2 \pi \kappa^2 (\omega_d)$ is the superconducting resonator leakage through the capacitors, and $\chi_{a_j}^{ A^{\rm in}}$ is  the linear response of the two-resonator system  driven by the input fields. The actual form of $\chi_{a_j}^{ A^{\rm in}}$ is rather cumbersome to give here, and can be found in  Appendix \ref{app:io}.
Putting all together, the transmitted signal for any of the three ports $j \lambda =1R, 2R, 2L$
is given by
\begin{equation}
T_{1L, j \lambda} = \frac{ \langle A^{\rm out}_{j, \lambda} \rangle }{
  \langle  A^{\rm in}_{1L}  \rangle }.
\end{equation}
The transmission with an input through any of the other ports can be
calculated in the same way.

In Fig. \ref{fig:ucs_trans}, we depict $T_{1L, 2R}$ as function of the
coupling and the input frequency.  As expected, the contour plot
resembles the real frequency plot in Fig. \ref{fig:ucs}(a).   The maximum in
the transmission coincides with the resonance frequencies.
Therefore,  in a transmission experiment the $\mathcal
PT$-symmetry  can be directly tested.

\section{Conclusions}

In this work we have shown that a circuit-QED architecture provides a flexible and highly versatile platform,
   with a small footprint, to explore the
physics of non-Hermitian $\mathcal {PT}$ systems.  Understanding that the latter is an effective theory, we have demonstrated how $\mathcal {PT}$ symmetry and its breaking emerges by
engineering a two resonator-two qubit Hamiltonian systems using tunable external drives, which is a natural strategy in circuit-QED systems.

This architecture has allowed us to probe the resonator-resonator interactions in various regimes of interaction strength thanks to the ability of achieving tunable coupling strength in circuit-QED, provided tunable gain and loss to delicately control the gain-loss ratio of the resonators, and opened the way to probe non-Hermitian dynamics in coupling regimes, ranging from weak to deep-coupling not only in the interaction between the resonators but in the interaction between the resonators and the qubits coupled to them for achieving gain and loss. We have shown that in the weak coupling regime of the resonator-resonator interaction, the $\mathcal{PT}$ symmetry is broken, i.e., the effective Hamiltonian exhibits non-real eigenvalues. 

By increasing the coupling, non-number-conserving terms start playing a significant role. This is the so-called \emph{ultrastrong} coupling regime that has already been explored experimentally in superconducting circuits \cite{Naether2014, Baust2015}.  In the $\mathcal {PT}$ scenario, this region corresponds to the unbroken (or the exact) $\mathcal {PT}$-symmetric phase. At much higher coupling strengths, the resonators become unstable. Crucially, this last transition is absent if we neglect the counter-rotating terms.  This regime corresponds to the \emph{deep-strong} coupling regime. More importantly, weak, strong, ultrastrong, and deep-strong regimes are differentiated by  transition points (either breaking symmetry or instability). We note that in previous studies (Rabi model \cite{Ashhab2007,Zueco2009, Casanova2010, Garziano2015}) where the qubit-resonator coupling was investigated in various coupling regimes, the borders between different regimes were diffuse. Revisiting those studies  by considering the $\mathcal {PT}$-symmetric resonator-resonator configuration proposed here may shed light on how different regimes in resonator-resonator coupling affect the quantum dynamics.

Finally, we have  shown  that the proposed circuit-QED architecture is experimentally accessible, no fine-tuning of the experimental parameters is necessary in order to observe the phenomenology imposed by  $\mathcal{PT}$ symmetry, and the basic concepts and applications that have been demonstrated in other platforms can be accessed and realized in this circuit-QED platform with a simple transmission experiment. We thus believe that this work will open the way to use circuit-QED as an ideal testbed to explore $\mathcal{PT}$-symmetric physics in
the quantum domain.


\begin{acknowledgements}
We would like to thank G\"oran Johansson for valuable discussions.
F.Q. acknowledges financial support from the Swedish Research Council and the Knut and Alice Wallenberg Foundation. 
S.K.O is supported by ARO Grant No. W911NF-18-1-0043 and by The Pennsylvania State University, Materials Research
Institute. F.N. is supported by the RIKEN iTHES Project, the MURI Center for Dynamic Magneto-Optics via the AFOSR award number FA9550-14-1-0040, the IMPACT program of JST, CREST.
D.Z. acknowledges support by the Spanish Ministerio de Economia y Competitividad within projects  MAT2014- 53432-C5-1-R and   FIS2014-55867 and  the Gobierno de Aragon (FENOL group).
F.Q., S.K.O. and D.Z. acknowledge the hospitality of Riken where part of this work was done.
\end{acknowledgements}


\bibliography{../PT_cQED}


\newpage
\begin{appendix}

\begin{widetext}



\section{Adiabatic elimination}
\label{app:adia}

Here we provide the details of the \emph {adiabatic elimination}. We consider a system (resonator) coupled to the environment through an ancillary element (qubit). The adiabatic elimination is rooted in the fact that the relaxation time scale of the ancilla is much faster than the typical time scale in which the system evolves [Cf. the hierarchy \ref{hiefull}].

\subsection{Single resonator case}

For the case of a single resonator, after driving the qubit and neglecting the rotating terms, we are left with the following Hamiltonian (in the interaction picture)
\begin{equation}\label{Heff1}
\widetilde{H}_c = g (b \sigma^+ + b^{\dagger} \sigma^- )
\qquad \qquad
b := G_- a + G_+
a^{\dagger}
\end{equation}
[Cf. Eq. (\ref{appHeffRWA}), for a single resonator $J=0$ and we consider $\delta=0$]. Coupling the qubit  to an environment, and assuming Markovianity, leads to the following non-unitary
evolution for the combined system of the resonator and the qubit (this is valid provided that $\epsilon \gg \gamma \gg g$)
\begin{equation}\label{QME_full}
{\rm d}_t \varrho = L_0[\varrho] + L_1[\varrho]
\end{equation}
where the state $\varrho$ lives in the total Hilbert space ($\mathcal{H}$) of the resonator ($\mathcal{H}_{\rm res}$) and the qubit ($\mathcal{H}_{\rm qub}$): $\mathcal{H} = \mathcal{H}_{\rm res} \otimes \mathcal{H}_{\rm qub}$. In addition, we have defined the Liouvillian operators:
\begin{equation}
L_0[\varrho] = \frac{\gamma}{2} (2 \sigma^- \varrho \sigma^+ - \lbrace \sigma^+ \sigma^- , \varrho \rbrace ) \,,
\end{equation}
and
\begin{equation}
L_1[\varrho] = -i[\widetilde{H}_c, \varrho ] .
\end{equation}
We will treat the $L_1$ part of the Liouvillian as a perturbation over $L_0$ (recall that $\gamma \gg g$). For this, we define the operator
\begin{equation}\label{rho_bar}
\bar{\varrho} = \exp \left( - L_0 t \right) \varrho
\end{equation}
which evolves in time according to
\begin{equation}\label{QMEL1}
{\rm d}_t \bar{\varrho} = \bar{L}_1 \bar{\varrho}
\end{equation}
with
\begin{equation}\label{L_1bar}
\bar{L}_1  = {\rm e}^{-L_0 t} L_1 {\rm e}^{L_0 t} .
\end{equation}
In order to deal with (\ref{QMEL1}) we will make use of projection operator techniques. The idea behind this method is to introduce two orthogonal projections, represented by the super operators $\mathcal{R}$ and $\mathcal{Q}$, with $\mathcal{R}^2 = \mathcal{R}$, $\mathcal{Q}^2 = \mathcal{Q}$, $\mathcal{RQ} = \mathcal{QR} = 0$ and $\mathcal{R} + \mathcal{Q} = \mathbb{I}$. This allows us to split the total density matrix $\varrho$ in a relevant part $\mu$ describing the resonator, and an irrelevant part describing the qubit $\varrho_{\rm qub}$. The action of $\mathcal{R}$ and $\mathcal{Q}$ on $\bar{\varrho}$ is defined by
\begin{eqnarray}
\mathcal{R}\bar{\varrho} &=& {\rm tr}_{\rm qub} (\bar{\varrho}) \otimes \varrho_{\rm qub} = \mu \otimes \varrho_{\rm qub} \\
\mathcal{Q}\bar{\varrho} &=& (\mathbb{I} - \mathcal{R}) \bar{\varrho}
\end{eqnarray}
Here $\varrho_{\rm qub}$ denotes some fixed state of the qubit. If we assume that the qubit undergoes a strongly dissipative dynamics, and in the absence of a pump, we can safely assume this state to be the ground state $\varrho_{\rm qub} = \vert \downarrow \rangle \langle \downarrow \vert$. As $L_0$ acts on the space of the qubit and $\mathcal{R}$ projects on the orthogonal space, these two super operators commute $[L_0, \mathcal{R}]=0$. This  guarantees that $\mathcal{R}\bar{\varrho} = \mathcal{R}\varrho$. Applying $\mathcal{R}$ and $\mathcal{Q}$ to (\ref{QMEL1}) we arrive at the following system
\begin{equation}\label{NZ1}
{\rm d}_t \mathcal{R}\bar{\varrho} = \mathcal{R} \bar{L}_1 \mathcal{R} \bar{\varrho} + \mathcal{R} \bar{L}_1 \mathcal{Q} \bar{\varrho}
\end{equation}
\begin{equation}\label{NZ2}
{\rm d}_t \mathcal{Q}\bar{\varrho} = \mathcal{Q} \bar{L}_1 \mathcal{Q} \bar{\varrho} + \mathcal{Q} \bar{L}_1 \mathcal{R} \bar{\varrho} .
\end{equation}
We first solve (\ref{NZ2})
\begin{equation}\label{NZ2_sol}
\mathcal{Q}\bar{\varrho}(t) = \mathcal{G}(t,0)\mathcal{Q}\bar{\varrho}(0) + \int_0^t {\rm d}s \, \mathcal{G}(t,s) \mathcal{Q} \bar{L}_1(s) \mathcal{R} \bar{\varrho}(s)
\end{equation}
where $\mathcal{G}$ denotes the time-ordered exponential
\begin{equation}
\mathcal{G}(t,s) = \mathcal{T} \exp \left( \int_s^t {\rm d}s' \mathcal{Q}\bar{L}_1(s') \right)
\end{equation}
which is the formal solution of
\begin{equation}
{\rm d}_t \mathcal{Q}\bar{\varrho} = \mathcal{Q} \bar{L}_1 \mathcal{Q} \bar{\varrho}
\end{equation}
being $\bar{L}_1$ time-dependent. We now substitute (\ref{NZ2_sol}) into (\ref{NZ1}) to obtain the so-called Nakajima-Zwanzig equation
\begin{equation}\label{NZeq}
{\rm d}_t \mathcal{R}\bar{\varrho} = \mathcal{R} \bar{L}_1 \mathcal{R} \bar{\varrho} + \mathcal{R} \bar{L}_1 \mathcal{G}(t,0) \mathcal{Q} \bar{\varrho}(0) + \mathcal{R} \bar{L}_1 \int_0^t {\rm d}s \, \mathcal{G}(t,s) \mathcal{Q} \bar{L}_1(s) \mathcal{R} \bar{\varrho}(s) .
\end{equation}
We can further simplify this equation as follows: from (\ref{rho_bar}) it follows that $\bar{\varrho}(0) = \varrho(0)$. If we assume an initial factorized state, the action of $\mathcal{R}$ on it equals to the action of the identity operator, and therefore $\mathcal{Q}\bar{\varrho}(0) = 0$. We now turn back to the equation for the state $\varrho$. From (\ref{rho_bar}) it follows
\begin{equation}
\mathcal{R}{\rm e}^{-L_0 t}{\rm d}_t \varrho = {\rm d}_t \mathcal{R}\bar{\varrho} + L_0 \mathcal{R} \bar{\varrho} .
\end{equation}
Replacing (\ref{NZeq}) in the former equation leads to
\begin{equation}
\mathcal{R}{\rm e}^{-L_0 t}{\rm d}_t \varrho = \mathcal{R} \bar{L}_1 \mathcal{R} \bar{\varrho} + \mathcal{R} \bar{L}_1 \int_0^t {\rm d}s \, \mathcal{G}(t,s) \mathcal{Q} \bar{L}_1(s) \mathcal{R} +  \mathcal{R} L_0 \mathcal{R} \bar{\varrho}(s) ,
\end{equation}
where we have made use of the fact that $\mathcal{RQ}\varrho =
\mathcal{R}\varrho - \mathcal{R}^2 \varrho = 0$. As usual, we will
assume  $\mathcal{R}L\mathcal{R}= 0$ \cite{Rivas2011} for the full Liouvillian $L = L_0 + L_1$. In our case this implies 
\begin{equation}
\mathcal{R} \bar{L}_1 \mathcal{R} \bar{\varrho} +  \mathcal{R} L_0 \mathcal{R} \bar{\varrho} = 0 .
\end{equation}
Thus, we are left with
\begin{equation}
\mathcal{R}{\rm e}^{-L_0 t}{\rm d}_t \varrho =  \mathcal{R} \bar{L}_1 \int_0^t {\rm d}s \, \mathcal{G}(t,s) \mathcal{Q} \bar{L}_1(s) \mathcal{R} .
\end{equation}
The lowest order expansion in the perturbation $L_1$ involves taking $\mathcal{G}(t,s) = \mathbb{I}$. This corresponds to second-order perturbation theory as $L_1$ already appears twice in the right-hand-side. Thus, from (\ref{L_1bar}), we finally have 
\begin{equation}
{\rm d}_t \mathcal{R} \varrho = \mathcal{R} L_1  \int_0^t {\rm d}s \, \exp \left[ L_0(t-s) \right] L_1  \mathcal{R} \varrho(s) .
\end{equation}
Tracing over the qubit we arrive at a  quantum master equation (QME) describing the effective dissipative dynamics of the resonator
\begin{equation}\label{QME_mu0}
{\rm d}_t \mu = -  \int_0^t {\rm d}s \, {\rm tr}_{\rm qub} [\widetilde{H}_c \,, \exp \left[ L_0(t-s) \right]  [\widetilde{H}_c \,, \mu(s) \otimes \varrho_{\rm qub}]] .
\end{equation}
We perform the following change of variables $s' = t-s$ and apply the Markov approximation, that is, $\mu(t-s') \rightarrow \mu(t)$. The final step consists of tracing out the qubit degrees of freedom. For this, we notice that $\sigma^{\pm}$ are eigen-operators of $L_0$ with eigenvalue $-\gamma /2$ ($L_0 \sigma^{\pm} = -(\gamma/2) \sigma^{\pm}$ ). Then, it is straightforward to show 
\begin{eqnarray}
[\widetilde{H}_c \,, {\rm e}^{L_0 s} [\widetilde{H}_c \,, \mu \otimes \varrho_{\rm qub}]] &=& \exp \left[-(\gamma/2) s \right] g^2 \left( [b^{\dagger},b\mu] - [b,\mu b^{\dagger}]  \right) \nonumber\\
&=& \exp \left[ -(\gamma /2)s \right] g^2 \left( - 2b\mu b^{\dagger} + \lbrace b^{\dagger}b, \mu \rbrace \right)
\end{eqnarray}
Then, equation (\ref{QME_mu0}) reduces to
\begin{equation}
{\rm d}_t \mu =  \int_0^t {\rm d}s \exp \left[ -(\gamma /2)s \right] g^2\left(  2b\mu b^{\dagger} - \lbrace b^{\dagger}b, \mu \rbrace \right) .
\end{equation}
Finally, integrating over time, in the limit $t \rightarrow \infty $ we obtain our desired result
\begin{equation}\label{QMEsingle}
{\rm d}_t \mu =  \frac{2 g^2}{\gamma}\left(  2b\mu b^{\dagger} - \lbrace b^{\dagger}b, \mu \rbrace \right) .
\end{equation}

The role of the $b$ operators is clear now. Using the drive on the auxiliary qubits, the effective dissipative dynamics on the resonator can have a non-trivial (not Gibbs) long-time dynamics. For example, whenever $G_-^2 - G_+^2 =1$, the $b$ operators become \emph{squeezed vacuum annihilator operators} and therefore, the stationary solution of Eq. (\ref{QMEsingle}) is a squeezed vacuum state.

\subsection{Coupled resonators}

Here we generalize the results derived for the single resonator case to  a chain of coupled cavities.
We consider one-dimensional, regular, and nearest-neighbor coupling between resonators in an array.
We consider two types of coupling.  The first is what we call as the RWA  coupling: $ \sim a_j^\dagger a_{j+1} + {\rm h.c.}$.  This conserves the total number of excitations in the lattice.

 The second type is called as the non-RWA coupling, which does not conserve the number of excitations:  $\sim (a_j + a_j^\dagger) (a_{j+1} + a_{j+1}^\dagger )$.
This appears naturally in dipole-dipole or displacement couplings in electromagnetic or mechanical systems.

Usually, the RWA coupling corresponds to the non-RWA in the weak-coupling regime.  
We emphasize that here both types of couplings are engineered. Therefore, it is not the interaction strength [always small - see \eqref{hiefull}] but the driving fields [Cf. Sects. \ref{sec:engine} and \ref{sec:JnRWA}] which dictate the type of coupling.

\subsubsection{RWA coupling}

We start by manipulating the coherent part of $H$ in Eq. (\ref{Hmicro}):
\begin{equation}
H = H_0 + H_c + H_{\rm drive}
\end{equation}
with
\begin{align}
H_0 &= \sum_j \left( \frac{\epsilon}{2}\sigma_j^z + \omega a_j^{\dagger} a_j \right)\\
H_c &= \sum_j   g\sigma_j^x (a_j^{\dagger}+a_j) +  J (a_j^{\dagger} +a_j)(a_{j+1}^{\dagger}+a_{j+1})\\
H_{\rm drive} & = \sum_j \sum_{\alpha} \lambda_{\alpha} \cos(\Omega_{\alpha}t)\sigma_j^z
\end{align}
where for simplicity, we assume that all resonators have the same frequency $\omega_r$ and all qubits have the same transition frequency $\epsilon$. Also, we assume that every resonator is coupled to its own qubit with the same strength $g$ and that the driving amplitudes are site-independent. Expanding the resonator operators in momentum space (plane wave basis), $a_k = N^{-1/2} \sum_j {\rm e}^{-i k j} a_j$, with $k \in 2 \pi / N \times \mathbb{Z}$, we can rewrite the total Hamiltonian as
\begin{equation}
H = H_0' + H_c' + H_{\rm drive}
\end{equation}
where
\begin{align}
H_0' & = \sum_j  \frac{\epsilon}{2}\sigma_j^z + \sum_k \varepsilon_k a_k^{\dagger} a_k \\
H_c' & = \sum_{k,j} g (e^{-ijk}\sigma_j^x a_k^{\dagger} + \mathrm{h.c.})
\end{align}
with $\varepsilon_k = \omega + 2 J \cos (k) $. The latter is valid whenever $\omega_r \gg J$. In this limit, the plane wave basis diagonalizes the inter-cavity interaction.
In the interaction picture with respect to $H_0' + H_{\rm drive}$, the interaction Hamiltonian is written as
\begin{equation}\label{Hint}
\widetilde{H}_c'(t) = \sum_{k,j} g\left\lbrace {\rm e}^{i j k} \left(   \sigma_j^+ {\rm e}^{2if(t)} + \sigma_j^- {\rm e}^{-2if(t)} \right) a_k {\rm e}^{-i\varepsilon_k t} + \mathrm{h.c.} \right\rbrace
\end{equation}
where the time-dependent term is given by
\begin{equation}
f(t)= \frac{\epsilon}{2} t + \sum_{\alpha} \frac{\lambda_{\alpha}}{\Omega_\alpha} \sin(\Omega_{\alpha}t)
\end{equation}
as can easily be obtained by integration. Now, we make use of the Jacobi-Anger expansion for the exponential terms
\begin{align}
\exp \left[ 2i f(t) \right] &= \exp \left[ i\left( \epsilon t + 2 \sum_{\alpha} \frac{\lambda_{\alpha}}{\Omega_{\alpha}} \sin( \Omega_{\alpha} t ) \right) \right] \\
&= \exp \left(i \epsilon t \right) \prod_{\alpha} \exp \left[ 2i \frac{\lambda_{\alpha}}{\Omega_{\alpha}} \sin( \Omega_{\alpha} t ) \right] \\
&= \exp \left(i \epsilon t \right) \prod_{\alpha} \sum_{n=-\infty}^{+\infty} J_n \left( 2\frac{\lambda_{\alpha}}{\Omega_{\lambda}} \right) \exp \left[ i n(\Omega_{\alpha}t ) \right] ,
\end{align}
where $J_n$ is the $n$-th Bessel function of the first kind. Up to  first order in the ratio $\lambda_{\alpha} / \Omega_{\alpha} \to 0$, we can safely neglect all orders of $J_n$, except $n=\pm1, \,0$. In addition, we select two driving frequencies ($\alpha = -, \,+$) $\Omega_- = \epsilon - (\omega + \delta)$ and $\Omega_+ = \epsilon + (\omega + \delta)$. According to our established hierarchy, $\epsilon \gg \omega$, we can neglect all the fast rotating terms. Therefore, we are left with
\begin{align}
\exp \left[ 2i f(t) \right] &= J_0\left( 2\frac{\lambda_-}{\Omega_-} \right) J_1\left( 2\frac{\lambda_+}{\Omega_+} \right) \exp \left[-i (\omega_r + \delta) t \right] + J_0\left( 2\frac{\lambda_+}{\Omega_+} \right) J_1\left( 2\frac{\lambda_-}{\Omega_-} \right) \exp \left[i (\omega_r + \delta) t \right] \,.
\end{align}
In order to simplify the notation, we will define
\begin{align}
G_+ &=  J_0\left( 2\frac{\lambda_+}{\Omega_+} \right) J_1\left( 2\frac{\lambda_-}{\Omega_-} \right) \\
G_- &=  J_0\left( 2\frac{\lambda_-}{\Omega_-} \right) J_1\left( 2\frac{\lambda_+}{\Omega_+} \right) .
\end{align}
Thus,
\begin{align}\label{exp_term}
 \exp \left[ 2i f(t) \right] &= G_- \exp \left[-i (\omega + \delta) t \right]  + G_+ \exp \left[i (\omega + \delta) t \right] \,.
\end{align}
Substituting (\ref{exp_term}) in (\ref{Hint}) yields
\begin{align}
\nonumber
\widetilde{H}_c'(t) &= \sum_{kj} g \left( G_+  {\rm e}^{ikj} \exp \left[ i(\omega - \epsilon_k + \delta)t \right] a_k + G_- {\rm e}^{-ikj} \exp \left[ -i(\omega - \epsilon_k + \delta)t \right] a_k^{\dagger} \right) \sigma^+_j + {\rm h.c.} \\
&= \sum_{kj} g \left( G_+  {\rm e}^{ikj} \exp \left[ -i(2J \cos(k) -\delta)t \right] a_k + G_- {\rm e}^{-ikj} \exp \left[i(2J \cos(k) -\delta)t \right] a_k^{\dagger} \right) \sigma^+_j + {\rm h.c.}
\end{align}
or
\begin{align}\label{Hint_final}
\widetilde{H}_c'(t) &= \sum_j g( b_j(t) \sigma_j^+ +  {\rm h.c.} )
\end{align}
with $b_j(t)$ given by
\begin{equation}\label{b_time}
b_j(t) = \sum_{k} \left( G_+  {\rm e}^{ikj} \exp \left[ -i(2J \cos(k)-\delta)t \right] a_k + G_- {\rm e}^{-ikj} \exp \left[i(2J \cos(k)-\delta)t \right] a_k^{\dagger} \right) .
\end{equation}

We will now proceed with the master equation
\begin{equation}\label{initial_step}
\dfrac{d\varrho}{dt} =  L_0 (\varrho) + L_1 (\varrho)  .
\end{equation}
Here $L_0$ describes the dissipation induced by the bath on the qubits  (recall that we only take into account spontaneous emission processes), therefore
\begin{equation}
L_0 (\varrho) = \sum_j \frac{\gamma}{2} (2 \sigma_j^- \varrho \sigma_j^+ - \sigma_j^+ \sigma_j^- \varrho - \varrho \sigma_j^+ \sigma_j^-)
\end{equation}
while $L_1$ describes the unitary evolution due to the coupling: $L_1 \varrho = -i[\widetilde{H}_c'(t),\varrho]$.

We want to study the dissipative dynamics induced on the resonators by the qubits. For a strong dissipative dynamics of the qubits, it is safe to assume that they remain fixed in the ground state. Therefore, we can adiabatically eliminate the degrees of freedom of the  qubits. We start by defining the projector $P$
\begin{equation}
P\varrho = \mu \otimes \varrho_{q,ss} = \mu \otimes \varrho_{q1,ss} \otimes ... \otimes \varrho_{qi,ss} ... \otimes \varrho_{qN,ss} .
\end{equation}
Here $\mu$ describes the system of resonators, and we take the ground state of all the qubits $\varrho_{qi,ss} = \vert \downarrow \rangle_{i i}\langle \downarrow \vert $ as a fixed state. In second order perturbation theory (in $L_1$) we obtain the following effective dynamics for the resonators
\begin{equation}\label{BornMarkov}
\dfrac{d\mu}{dt} = - \int_0^{\infty} d\tau \, \mathrm{Tr_q}[\widetilde{H}_c'(t),e^{L_0 \tau}([\widetilde{H}_c'(t-\tau),\mu(t)\otimes \varrho_{q,ss}])]
\end{equation}
where the Born-Markov approximation has already been performed. Expanding the commutators in (\ref{BornMarkov}) we can perform the partial trace over the qubits. For this, we must take into account (\ref{Hint_final}) and that $\sigma_j^+$ and $\sigma_j^-$ are eigenstates of the super-operator $L_0$ both with eigenvalue $-\gamma / 2$. Thus, we are left with
\begin{equation}\label{BornMarkov2}
\dfrac{d\mu}{dt} =  \int_0^{\infty} d\tau \, {\rm e}^{-\gamma/2 \tau} \sum_j g^2 \left(  [b_j^{\dagger}(t), b_j(t-\tau) \mu(t)] - [b_j(t), \mu(t) b_j^{\dagger}(t-\tau)] \right) .
\end{equation}
Expanding the $b_j(t)$ operators and performing the integration over the variable $\tau$ yields
\begin{equation}
\int_0^{\infty} {\rm d}\tau \, \exp \left[ -(\gamma /2 \pm i (2J\cos(k) + \delta))\tau \right] = \frac{2/ \gamma}{1 \pm 2i (2J\cos(k)+\delta) /\gamma}  = \frac{2}{ \gamma}
\end{equation}
which follows from the hierarchy of energies considered in this work [Cf. Eq. \eqref{hiefull}].
From this, we arrive to the QME in the interaction picture in position space
\begin{align}\label{QME_position}
\dfrac{d\mu}{dt} =  \frac{2 g^2}{\gamma} \sum_j  2 b_j(t) \mu(t) b_j^{\dagger}(t) - \lbrace b_j^{\dagger}(t)b_j(t) ,\mu(t) \rbrace .
\end{align}
We note that $\gamma \gg J$ is required to arrive at (\ref{QME_position}). In fact, the time evolution of the $b_j$ operators could be more intricate, \emph{i.e.}, non-reducible to an analytic expression. For a general evolution operator $U$, the time evolution is given by $b_j(t) = U^{\dagger}(t) b_j(0) U(t)$. Assuming a time-independent Hamiltonian $H$, we have $U(t) = {\rm exp}(-iHt)$. Decomposing the latter into eigenstates of $H$ ($H \vert \alpha \rangle = E_{\alpha} \vert \alpha \rangle $) leads to $U(t) = \sum_{\alpha}{\rm exp}(-iE_{\alpha}t) \vert \alpha \rangle \langle \alpha \vert$.
Performing the time integration in (\ref{BornMarkov2}) will lead to terms of the form
\begin{align}
\int_0^{\infty} {\rm d}\tau \, {\rm e}^{-\gamma/2 \tau} b_j(t-\tau) &= \sum_{\alpha \beta} {\rm e}^{i(E_{\beta} - E_{\alpha})t} \int_0^{\infty} {\rm d}\tau \, {\rm e}^{-(\gamma/2 - i(E_{\alpha} - E_{\beta}) )\tau} \vert \beta \rangle \langle \beta \vert b_j(0) \vert \alpha \rangle \langle \alpha \vert .
\end{align}
If the characteristic energies associated to $H$ are much more smaller than the coupling to the environment ($\gamma \gg E_{\alpha}$), we have again for the integral
\begin{equation}
\int_0^{\infty} {\rm d}\tau \, \exp \left[-(\gamma /2 - i (E_{\alpha} - E_{\beta}))\tau \right] = \frac{2/ \gamma}{1 - 2i (E_{\alpha} - E_{\beta}) /\gamma}  = \frac{2}{ \gamma} .
\end{equation}
Therefore,
\begin{align}
\nonumber
\int_0^{\infty} {\rm d}\tau \, \exp \left( -\gamma/2 \tau \right) b_j(t-\tau) &= \frac{2}{\gamma} \sum_{\alpha \beta} {\rm e}^{i(E_{\beta} - E_{\alpha})t} \vert \beta \rangle \langle \beta \vert b_j(0) \vert \alpha \rangle \langle \alpha \vert \\
&= \frac{2}{\gamma} b_j(t)
\end{align}
and similarly for the Hermitian conjugated terms. This leads again to Eq. (\ref{QME_position}).

The final step consists of going back to the Schr\"odinger picture and express the QME in momentum space. From (\ref{b_time}), the momentum space operators $a_k$ evolve in time according to $a_k \,{\rm exp}[-i(2J \cos (k) - \delta)t]$. Going back to the Schr\"odinger picture implies canceling out these rotating terms. We can fulfil this condition by applying the following transformation
\begin{equation}
a_k = U_0(t) a_k(t) U_0^\dagger(t)
\end{equation}
with $ U_0(t) = {\rm exp}\{ -i [2J \cos (k) - \delta ] t \}$,
to both sides of (\ref{QME_position}). Doing so, we arrive to the desired result
\begin{align}
\label{QME_momentum_final}
\frac{d\mu}{dt}=\sum_k
-i \omega_k
[ a_k^{\dagger}a_k,\mu]
+ \frac{2 g^2}{\gamma}(2b_k\varrho
b_k^{\dagger}-\{b_k^{\dagger} b_k,\mu \})
\end{align}
where we have defined $\omega_k = 2J \cos (k) - \delta$, and $b_k = G_+ a_k + G_- a_{-k}^\dagger $.

\subsubsection{Non-RWA coupling}
Following Sect. \ref{sec:JnRWA}, after engineering $H$ by means of a two-color driving and  a time-dependent coupling $J(t)$, we arrive to the effective Hamiltonian
\begin{equation}
H_{\rm eff}^{\rm nRWA} = -\sum_j \delta a_j^{\dagger} a_j + J(a_j^{\dagger} +
a_j)(a_{j+1}^{\dagger} + a_{j+1}) + \sum_j g \left( G_{j+} \sigma^+_j a_j + G_{j-} \sigma^+_j a_j^{\dagger} + {\rm h.c.} \right) .
\end{equation}
In order to obtain the dissipative dynamics for the resonators, we proceed as we did in the previous section.
However, instead of moving to the momentum space, we rewrite $H_{\rm eff}^{\rm nRWA}$ in an interaction picture with respect to $\bar{H} = -\sum_j \delta a_j^{\dagger} a_j +J(a_j^{\dagger} +
a_j)(a_{j+1}^{\dagger} + a_{j+1})$. This leads to
\begin{equation}
H_{\rm eff}^{\rm nRWA}(t) = \sum_j g \left( G_{j+} a_j(t) \sigma^+_j + G_{j-} a_j^{\dagger}(t) \sigma^+_j + {\rm h.c.} \right)
\end{equation}
where $a_j(t) = {\rm e}^{i
  \bar H t} a_j {\rm e}^{-i
  \bar H t} $.
By  introducing the operators $b_j(t) = G_{j+} a_j(t) + G_{j-}
a_j^{\dagger}(t) $, the above yields
\begin{equation}
H_{\rm eff}^{\rm nRWA}(t) = \sum_j g( b_j(t) \sigma^+_j + {\rm h.c.} ) .
\end{equation}
Here,  we do not know the explicit time dependence of the $b_j$ operators. Assuming again a strong dissipation for the qubits, we can follow the general steps in Eqs. (\ref{initial_step}) -
(\ref{BornMarkov2}). Regardless of the explicit evolution of the $b_j$'s, whenever the energy scale associated with the transformation Hamiltonian (in this case $\bar{H}$) is much more smaller than $\gamma$ (as dictated by \ref{hiefull}), we can always reduce (\ref{BornMarkov2}) to (\ref{QME_position}). Going back to the Schr\"odinger picture we obtain,
\begin{align}
\dfrac{d\mu}{dt} = -i \left[\sum_j -\delta a_j^{\dagger} a_j + J(a_1^{\dagger} + a_1)(a_2^{\dagger} + a_2),\mu \right] + \frac{2 g^2}{\gamma} \sum_j  2 b_j \mu b_j^{\dagger} - \lbrace b_j^{\dagger}b_j ,\mu \rbrace .
\end{align}


\section{Eigenvalues for the case of unbalanced gain and loss}
\label{app:ibev}

\begin{figure}[t]
\begin{center}
\includegraphics[width=0.5\columnwidth]{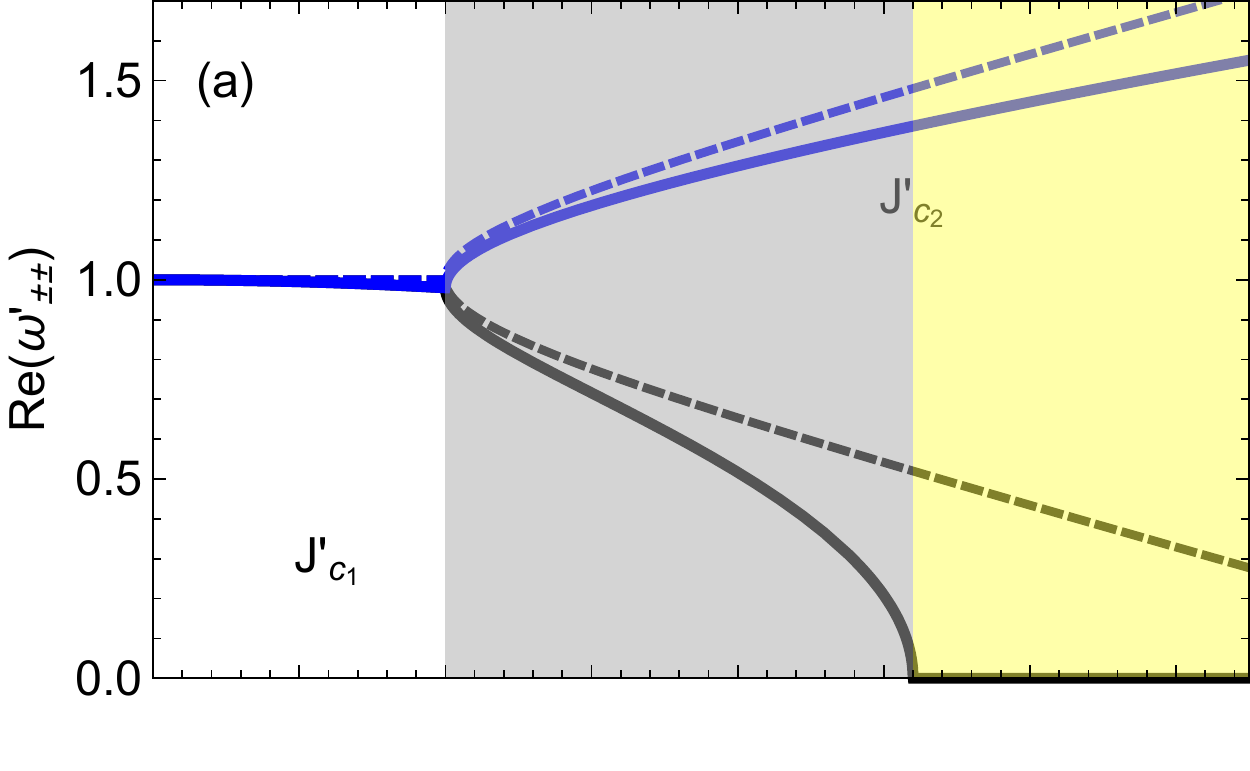}
\includegraphics[width=0.5\columnwidth]{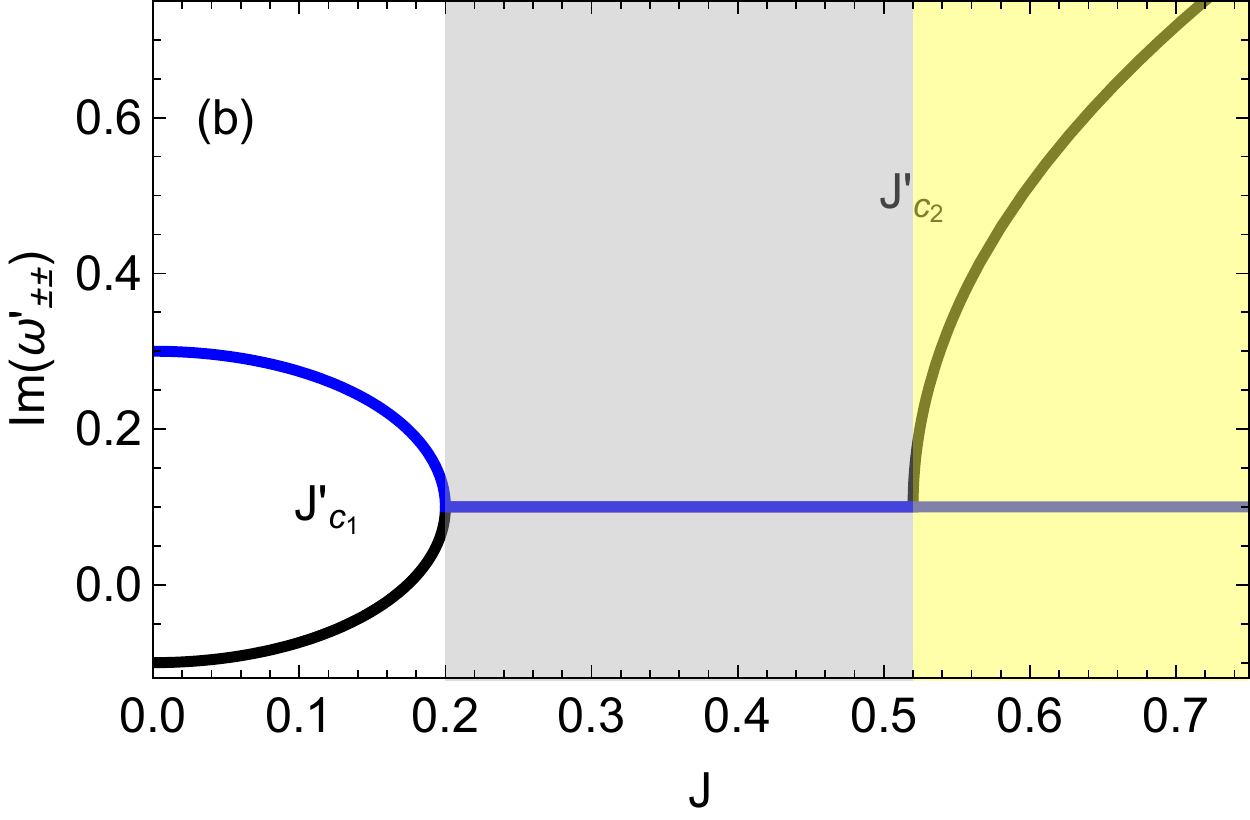}
\end{center}
\caption{   (Color online)
Eigenfrequencies for the case of unbalanced gain and loss between the resonators \eqref{omegas_imb}.
The background colors distinguish among three regions:
$J<J_{c_1}'$ (white),
  $J_{c_1}'<J<J_{c_2}'$(light gray) and $J_{c_2}'<J$ (light yellow) [Cf. Eqs. \eqref{Jc1imb} and \eqref{Jc2imb}].
The
eigenfrequency  $\omega_{+-}'$ is plotted in  black while
$\omega_{++}'$ in
 blue. The dashed lines correspond
 to the RWA \eqref{appHeffRWA} while the solid
 lines correspond to the most general case \eqref{Heffus}.
In (a) we plot the real part of $\omega_{+ \pm}'$ and in (b) the
imaginary part.  The parameters chosen are,
$\widetilde \gamma_1= 0.1$, $\widetilde \gamma_2=0.3$, and $\delta= 1$.
}
\label{fig:imbalanced}
\end{figure}

Diagonalizing  Eq.~\eqref{alphaM}  for the general (i.e., unbalanced gain and loss)
case $\tilde{\gamma}_1\neq\tilde{\gamma}_2$,  we obtain the eigenfrequencies
\begin{align}
\label{omegas_imb}
&&\omega'_{ \pm \pm}=
\pm
\left[
\delta^2  - \frac{(\widetilde \gamma_1 + \widetilde \gamma_2)^2}{4}
\pm 2 \delta \left( J^2 - \frac{(\widetilde \gamma_1 + \widetilde \gamma_2)^2}{4} \right)^{1/2} \right]^{1/2}
\,
+ i \frac{\widetilde \gamma_2 - \widetilde \gamma_1}{2}
\, .
\end{align}
It is seen that, whenever $\widetilde{\gamma}_1\neq\widetilde{\gamma}_2$, there is no region in the parameter space where \eqref{omegas_imb}  is real (i.e., no complex part). However, the eigenvalues may or may not coincide in their real or imaginary parts depending on the square roots in \eqref{omegas_imb}. The generalizations of the critical values \eqref{Jc1} and \eqref{Jc2} to the imbalanced case therefore are:

\begin{equation}
\label{Jc1imb}
J < J_{c1}' =  \frac{\widetilde \gamma_1 + \widetilde \gamma_2}{2}
\, ,
\end{equation}
and,
\begin{equation}
\label{Jc2imb}
J_{c2}' = \frac{(\widetilde \gamma_1+ \widetilde \gamma_2)^2+4 \delta ^2}{8 \delta } < J.
\end{equation}

This general behavior, for both RWA and non-RWA cases, is shown in Fig. \ref{fig:imbalanced}. There is always a non-zero imaginary contribution to the normal frequencies [Cf. last term in
\eqref{omegas_imb}].  Even in this case, the phase transitions are clearly seen. 


\section{Input-Output}
\label{app:io}

In a fully quantum treatment, the system (in our case, the two resonator -two qubit circuit) and the input-output transmission lines
 [Cf. Fig. \ref{fig:scheme}]  are formally described by the Hamiltonian
\begin{equation}
H = H_{\rm sys} + \sum_{j, \lambda} \int {\rm d} \omega A^\dagger_{j,
  \lambda} (\omega) A_{j,
  \lambda} (\omega) + \sum_j \kappa_{j, \lambda} (\omega)
 \Big ( a_j   A^\dagger_{j,
  \lambda} (\omega) +   {\rm h.c.} \Big )
\end{equation}
with $[A_{j, \lambda} (\omega), A_{j^\prime, \lambda^\prime}
(\omega^\prime)^\dagger ] = \delta_{j, j^\prime} \, \delta_{\lambda,
  \lambda^\prime} \, \delta (\omega - \omega^\prime)$ the quantized
modes in the input lines.
Here we use the same labeling $j=1,2$, and $\lambda= L, R$, for different ports, as described in the main text.
For the sake of simplicity, we will consider
that both resonators are equally and symmetrically coupled to the corresponding transmission lines.

The Heisenberg equations for the operators $A_{j, \lambda} (\omega)$ together with the definitions:
\begin{align}
A^{\rm in}_{j, \lambda}(t) &= \int_0^\infty {\rm
  d}\omega / \sqrt{2 \pi} \;  A_{j, \lambda} (\omega, t_0)  \, {\rm e}^{-i \omega
  t}
\\
A^{\rm out}_{j, \lambda} (t) &= \int_0^\infty {\rm
  d}\omega / \sqrt{2 \pi} \; A_{j, \lambda} (\omega, t_f)  \, {\rm e}^{-i \omega
  t}
\; ,
\end{align}
yield the following relation for the output and input fields:
\begin{equation}
A^{\rm out}_{j, \lambda} (t) = A^{\rm in}_{j, \lambda}(t)
- i \int_0^\infty \frac{{\rm d}\omega}{\sqrt{2 \pi}} {\rm e}^{- i
  \omega t} \kappa (\omega) \int_{-\infty}^\infty {\rm d} \tau {\rm e}^{i
  \omega \tau} a_j  (\tau)
\end{equation}
where $a_j (\tau) = {\rm e}^{i H \tau} a_j {\rm
  e}^{-i H \tau}$.
\

In order to calculate the evolution of  $a_j (\tau) $,  we assume coherent and low-power input signals. This allows us to write
the evolution for $a_j$ analogous to Eqs. (\ref{aeqio}).
\begin{equation}
\label{aeqio-app}
\frac{d}{d t} \langle  a_j   \rangle
=
i \langle [H_{\rm eff}, a_{j, \lambda}] \rangle
-i \sum_\lambda \langle A^{\rm in}_{j, \lambda} (t) \rangle
+\frac{2}{\gamma}
\langle D^\dagger [b_j] a_j \rangle
+
2 \kappa  \langle D^\dagger [a_j] a_j \rangle
\; .
\end{equation}
As compared to \eqref{aeqio}, the above expression includes an extra dissipation channel (always loss) due to the coupling to the feed lines, plus the
driving due to the input signal. Without loss of generality we can work with a monochromatic signal (any signal can be written in terms of monochromatic ones)
\begin{equation}
\label{A-input}
 \langle A^{\rm in}_{1,L} (t) \rangle = \alpha \exp \left(i \omega_d t \right) .
\end{equation}
Using Linear Response Theory (LRT) we find
\begin{equation}
\label{chi}
\langle a_j (\tau) \rangle = \chi_{a_j}^{A^{\rm in}}  \exp(- i
  \omega_d t)
\qquad
\quad
\chi_{a_j}^{A^{\rm in}} = \frac{1}{\alpha} \left (
\Delta a_r (0) - i \omega \int_0^\infty {\rm d} t \Delta a_r (t) \exp(- i \omega t)
\right )
\end{equation}
where $\Delta a_r (t) = \langle a_j (t) \rangle -  \langle a_j (t \to \infty) \rangle  $ is the so-called relaxation response evolving
with \eqref{aeqio}, without the drive.  The initial condition $\langle
a_j (0) \rangle$ is the equilibrium solution of \eqref{aeqio} with
a constant drive $\langle A^{\rm in}_{1 R} (t)
\rangle= \alpha$.  This is the well known LRT result where the AC-response can be related to a DC relaxation experiment. Within this result at hand, the calculation for $\langle a_j (\tau)
\rangle$ is reduced to solve, in our case, a linear set of four coupled differential equations.

\subsection{The RWA case}

Here we will discuss some simplifications that can be used for computing $\langle a_j\rangle$ in the RWA case.
The triumph of LRT is to avoid the time dependent problem \eqref{aeqio} in the AC-response by using the formula in Eq. \eqref{chi}.
However, in the RWA case, we do not need to use this general formalism.   The calculations are simpler by noting that we can work in a rotating basis with the
drive, $\omega_d$ [Cf. Eq. \eqref{A-input}]. In Eq.~\eqref{aeqio-app}, we have the terms  $a_j$  and $a_{j^\prime}^\dagger$ appearing together.  Thus, the equations are time independent with just a  shift in the frequency, $\delta - \omega_d$. Introducing the notation
\begin{equation}
\alpha_j := \langle a_j  \rangle
\end{equation}
we can write
\begin{align}
\label{PT}
i{\rm d}_t \alpha_1 &=  (\delta - \omega_d) \alpha_1 + J \alpha_2 - i(\widetilde \gamma_1 +
\kappa )\alpha_1 - i \sqrt{\kappa} \langle A^{\rm in}_{1,L} \rangle \\ \nonumber
i{\rm d}_t \alpha_2 &=   (\delta - \omega_d) \alpha_2 + J \alpha_1 +
                      i(\widetilde \gamma_2 - \kappa ) \alpha_2
\; .
\end{align}
There is an analogous set for the complex conjugates. Let us, without loss of generality, consider the case where
the input is sent through the port $1L$.  Then we have the following matrix form:
\begin{equation}
\label{PT-matrix}
i{\rm d}_t  {\bf \alpha} = (M_{\rm RWA} -i \kappa {\mathbb I}_2 ){\bf
  \alpha} + {\bf j}
\end{equation}
with,
\begin{equation}
{\bf j}= - i \sqrt{\kappa} \left(
\begin{array}{cc}
\langle A^{\rm in}_{1 L} \rangle
\\
0
\end{array}
\right )
\end{equation}

In the rotating basis, the coherent input state \eqref{A-input} is $\langle A^{\rm in}_{1 L} \rangle = \alpha$. Then we find the response function,
\begin{equation}
\chi_{a_j}^{A^{\rm in}} = \alpha^{\rm eq}_j
\end{equation}
with
\begin{eqnarray}
\label{alpha1}
\alpha_1^{\rm eq} &=& \frac{i\sqrt{\kappa}\alpha}{\omega_+ \omega_-} \left[
  \delta + i\left( \widetilde \gamma_2 - \kappa \right) \right] \\ \label{alpha2}
\alpha_2^{\rm eq} &=& -\frac{i\sqrt{\kappa}\alpha}{\omega_+ \omega_-}J
\end{eqnarray}
where $\omega_{\pm}$ is given by
\begin{equation}\label{eigen_2}
\omega_{\pm}= \delta+\frac{1}{2} \left[ i (\widetilde \gamma_2- \widetilde \gamma_1 - 2\kappa) \pm \sqrt{ 4 J^2-(\widetilde \gamma_1+\widetilde \gamma_2)^2 } \right].
\end{equation}
Then it is not difficult to solve the \emph{input-output} relations in Eq. \eqref{in-out}. For example, we can measure the transmitted signal in port $2L$ or
$2R$ obtaining
\begin{equation}\label{out_lower_right}
\langle A_{2 R}^{\rm out} \rangle = \frac{i\kappa \epsilon J}{\omega_+ \omega_-}
\end{equation}

\end{widetext}

\end{appendix}

\end{document}